\documentclass[a4paper,11pt]{article}
\pdfoutput=1 

\usepackage{jcappub} 
\usepackage[toc,page]{appendix}
\usepackage{subcaption}
\usepackage{amsmath}

\usepackage{graphicx}                
\usepackage{multirow}
\usepackage[T1]{fontenc} 
\usepackage[table]{xcolor}
\usepackage{float}
\def\noi{{\noindent}}

\definecolor{verde}{rgb}{0,0.5,0}

\def\be{\begin{equation}}
\def\ee{\end{equation}}
\def\bea{\begin{eqnarray}}
\def\eea{\end{eqnarray}}
\def\be{\begin{equation}}
\def\ee{\end{equation}}
\def\ba{\begin{eqnarray}}
\def\ea{\end{eqnarray}}

\hypersetup{pdftitle={Novel CMB constraints on alpha in alpha-attractor models of inflation}
}
\usepackage{bm}

\title{\boldmath 
Novel CMB constraints on the $\alpha$ parameter in alpha-attractor models}

\author[a,b]{Laura Iacconi,}
\author[c,b]{Matteo Fasiello,}
\author[d,e]{Jussi V\"{a}liviita}
\author[b]{and David Wands}
\affiliation{$^{a}$Astronomy Unit, Queen Mary University of London, \\ Mile End Road, London, E1 4NS, U.K.}
\affiliation{$^{b}$ Institute of Cosmology \& Gravitation, University of Portsmouth,\\ Burnaby Road, Portsmouth, PO1 3FX, U.K.}
\affiliation{$^{c}$ Instituto de F\'{i}sica T\'{e}orica UAM-CSIC,\\ Calle Nicol\'{a}s Cabrera 13-15, 28049, Madrid, Spain}
\affiliation{$^{d}$ Department of Physics, University of Helsinki, \\ P.O. Box 64, FIN-00014, Finland}
\affiliation{$^{e}$ Helsinki Institute of Physics, University of Helsinki,\\ Gustaf H{\"a}llstr{\"o}min katu 2,  Helsinki, Finland}

\emailAdd{l.iacconi@qmul.ac.uk}
\emailAdd{matteo.fasiello@csic.es}
\emailAdd{jussi.valiviita@helsinki.fi}
\emailAdd{david.wands@port.ac.uk}

\abstract{Cosmological $\alpha$-attractors are a compelling class of inflationary models. They lead to universal predictions for large-scale observables, broadly independent from the functional form of the inflaton potential. In this work we derive \textit{improved} analytical predictions for the large-scale observables, whose dependence on the duration of reheating and the parameter $\alpha$ is made explicit. 
We compare these with \textit{Planck} and BICEP/Keck 2018 data in the framework of a Bayesian study, employing uniform logarithmic and linear priors for $\alpha$. Our improved universal predictions allow direct constraints on the duration of reheating. Furthermore, while it is well-known that CMB constraints on the tensor-to-scalar ratio can be used to place an upper bound on the $\alpha$ parameter, we demonstrate that including the $\alpha$-dependence of the scalar spectral tilt yields novel constraints on $\alpha$. 
In particular, for small $\alpha$, the scalar spectral tilt scales with $\log_{10}\alpha$, 
regardless of the specific potential shape. For decreasing $\alpha$, this eventually puts the models in tension with CMB measurements, bounding the magnitude of $\alpha$ from below. Therefore, in addition to the upper bound from the tensor-to-scalar ratio, we derive the first lower bound on the magnitude of $\alpha$ for $\alpha$-attractor T-models, $\log_{10}{\alpha} = -4.2^{+5.4}_{-8.6}$ at $95\%$ C.L. .}

\begin{document}
	\maketitle
	\flushbottom
	
\section{Introduction}
\label{sec: intro}

\noi Cosmological inflation was originally proposed \cite{Guth:1980zm, Linde:1981mu, Starobinsky:1980te,Albrecht:1982wi} to solve issues related to the initial conditions in the standard homogeneous hot big bang model of cosmology. Shortly after the initial proposal, it was realised that, remarkably, it also provides the quantum seeds for the large-scale structure of the cosmic web~\cite{Mukhanov:1981xt, Starobinsky:1982ee, Guth:1982ec, Hawking:1982cz, Bardeen:1983qw}. As a result of this early success and the continued agreement with observations, it now constitutes the main paradigm in the description of the very early universe.

Cosmological $\alpha$--attractors \cite{Kallosh:2013hoa, Kallosh:2013daa, Ferrara:2013rsa, Kallosh:2013pby, Kallosh:2013lkr,Kallosh:2013maa, Kallosh:2013tua, Kallosh:2013yoa} are a class of inflationary models. They provide a compelling mechanism for acceleration in the early universe for several reasons. They lead to universal predictions for large-scale observables that are broadly independent of the details of the inflationary potential \cite{Kallosh:2013hoa}. The class of $\alpha$--attractor models can be embedded in supergravity theories and M-theory (see e.g.\cite{Ferrara:2013rsa}), and their predictions lie close to the center of current observational bounds on the primordial power spectra \cite{Planck:2018jri}.

We shall focus on simple $\alpha$-attractor T-models \cite{Kallosh:2013hoa}, where the inflaton potential is given by 
\begin{equation}
\label{single field potential}
    V(\phi)=V_0\tanh^p{\left( \frac{\phi}{\sqrt{6\alpha}}\right)} \;,
\end{equation} 
where $p$ is an even integer $p\geq 2$ and we have set the reduced Planck mass, $M_\text{Pl}\equiv(8\pi G_N)^{-1/2}$, to unity.

Predictions for the large-scale observables of cosmological $\alpha$-attractors can be estimated analytically within the slow-roll approximation.
These depend primarily on ${\Delta N_\text{CMB}}$, the number of e-foldings between horizon crossing of the CMB scale during inflation ($k_\text{CMB}=aH$, where $k_\text{CMB}$ is usually identified with the scale $0.05\,\text{Mpc}^{-1}$) and the end of inflation, plus small corrections from the model parameters, e.g. $\alpha$ and $p$ in the case of T-models \eqref{single field potential}. For the scalar power spectrum spectral tilt, $n_s$, and the tensor-to-scalar ratio, $r$, these are \cite{Kallosh:2013yoa, Kallosh:2013hoa}
\begin{align}
\label{ns p dependence}
    n_s& \approx 1-\frac{2\Delta N_\text{CMB}+\frac{1}{p}\sqrt{3\alpha(3\alpha+p^2)}+\frac{3\alpha}{2}}{\Delta N_\text{CMB}^2 +\frac{\Delta N_\text{CMB}}{p}\sqrt{3\alpha(3\alpha+p^2)}+\frac{3\alpha}{4}} \;, \\
\label{r p dependence}
    r& \approx \frac{12\alpha}{\Delta N_\text{CMB}^2 +\frac{\Delta N_\text{CMB}}{p}\sqrt{3\alpha(3\alpha+p^2)}+\frac{3\alpha}{4}} \;. 
\end{align}
In the large $\Delta N_\text{CMB}$ limit and at leading order in the ${\Delta N_\text{CMB}}^{-1}$ expansion, the expressions above can be simplified to give \cite{Kallosh:2013hoa, Kallosh:2013yoa}
\begin{align}
\label{ns universal}
    n_s & \approx 1-\frac{2}{\Delta N_\text{CMB}}\;, \\
\label{r universal}
    r & \approx \frac{12\alpha}{{\Delta N_\text{CMB}}^2}  \;.
\end{align}
Remarkably, in this limit the predictions become \textit{universal}, i.e. they do not explicitly depend on the details of the inflaton potential. The only quantity determining the spectral tilt is $\Delta N_\text{CMB}$, while $r$ also explicitly depends on the 
parameter $\alpha$. For $50\lesssim \Delta N_\text{CMB} \lesssim 60$ and $\alpha \lesssim \mathcal{O}(1)$, the predictions \eqref{ns universal} and \eqref{r universal} sit comfortably within the bounds from current CMB observations. In order to distinguish between the predictions in Eqs.\eqref{ns p dependence}-\eqref{r p dependence} and those in Eqs.\eqref{ns universal}-\eqref{r universal}, we shall refer to the former as the \textit{extended} predictions and to the latter as the \textit{standard universal} predictions. 

Crucially however, $\Delta N_\text{CMB}$ does implicitly depend on the inflaton potential, and therefore on the model parameters. Understanding this dependence allows us to explore the detailed behavior of the large-scale observables with respect to changes in the model parameters.
In this work we show that including the implicit $\alpha$-dependence within a Bayesian analysis yields novel constraints on the $\alpha$ parameter, enabling new tests of $\alpha$-attractors with existing CMB data. 

\medskip
\textit{Content:} In this work we will consider $\alpha$-attractor T-models \eqref{single field potential}, and explore cases with $p=2$ and $p=4$. We introduce the T-models in section \ref{sec: inflation and reheating}, where we also discuss the reheating stage following the end of inflation. In section \ref{sec:alpha scaling of the CMB observables} we numerically and analytically explore  the dependence of large-scale observables on $\alpha$ and on the duration of reheating for $p=2$. In particular, in section \ref{sec: analytic Delta N CMB} we derive an analytic expression for $\Delta N_\text{CMB}$, making the dependence on $\alpha$ and on the duration of reheating explicit. In section \ref{sec:improved predictions} we use this analytic result to improve the extended predictions, Eqs.\eqref{ns p dependence}-\eqref{r p dependence}, and the
standard universal predictions, Eq.\eqref{ns universal}-\eqref{r universal}, and compare these with the numerical large-scale observables. In section \ref{sec:CMB bounds} we compare our improved predictions with current \textit{Planck} \cite{Planck:2018nkj, Planck:2019nip} and BICEP/Keck \cite{ BICEPKeck:2021gln} data in the framework of Bayesian statistics, and derive posterior distributions for the inflationary parameters. 
In section \ref{sec:models with p=4} we consider models with $p=4$ and establish that the $\alpha$-dependence of  large-scale observables and the related results for $p=2$, in sections \ref{sec:alpha scaling of the CMB observables} and \ref{sec:CMB bounds}, generalise to a different choice of $p$. We discuss our results in section \ref{sec: discussion} and provide additional material in the appendices. In appendix \ref{app: inflaton oscillations} we compare the inflaton oscillations following the end of inflation, and the effective equation of state parameter during this phase, for models with large and small $\alpha$. In appendix \ref{app:slow-roll observables} we describe in more detail the numerical predictions for $n_s$ and $r$. 

\medskip
\textit{Conventions:} Throughout this work, we consider a spatially-flat Friedmann--Lema\^{i}tre--Robertson--Walker universe, with line element $\text{d}s^2=-\text{d}t^2+a^2(t)\delta_{ij}\text{d}x^i\text{d}x^j$, where $t$ denotes cosmic time and $a(t)$ is the scale factor. The Hubble rate is defined as $H\equiv {\dot a}/{a}$,  where a derivative with respect to cosmic time is denoted by $\dot f \equiv {\mathrm{d}f}/{\mathrm{d}t}$. The number of e-folds of expansion is defined as $N\equiv \int\,H(t)\mathrm{d}t$ and $f'\equiv {\mathrm{d}f}/{\mathrm{d}N}$. We use natural units and set the reduced Planck mass, $M_\text{Pl}\equiv(8\pi G_N)^{-1/2}$, to unity unless otherwise stated.

\section{Cosmological inflation and reheating}
\label{sec: inflation and reheating}
Cosmological $\alpha$-attractors are naturally implemented in supergravity and are usually formulated in terms of a complex field, $Z$, belonging to the Poincaré hyperbolic disc \cite{Carrasco:2015uma, Kallosh:2015zsa}, characterised by $|Z|\leq1$ and radius $R=\sqrt{3\alpha}$. The complex field $Z$ is endowed with potential energy $V(Z,\,\bar Z)$, regular everywhere on the disk. The corresponding Lagrangian reads
\begin{equation}
    \label{kinetic lagrangian 1}
    \mathcal{L}=\frac{1}{2}R -3\alpha\frac{\partial_\mu Z\partial^\mu\bar Z}{\left(1-Z\bar Z \right)^2} -V(Z, \, \bar Z) \;.
\end{equation}
The potential $V(Z,\,\bar Z)$ is a generic function of $Z$ and $\bar Z$. In this work we focus on $\alpha$-attractor T-models \cite{Kallosh:2015zsa} and consider contributions to $V(Z,\,\bar Z)$ invariant under a phase-shift $Z\to e^{i\varphi}Z$,
\begin{equation}
\label{V of Z}
    V(Z,\,\bar Z)\equiv \sum_{i=1}^\infty  c_{2i} \,(Z\bar Z)^{i} = c_2\, Z \bar Z + c_4\,  (Z \bar Z)^2 + \cdots \;.
\end{equation}
The complex field $Z$ can be parametrised in terms of two scalar fields,  the radial and angular fields ($r$ and $\theta$ respectively),
\begin{equation}
\label{canonical field}
    Z\equiv r e^{i\theta} \equiv \tanh{\left(\frac{\phi}{\sqrt{6\alpha}}\right)} e^{i\theta} \;,
\end{equation}
where in the last step we have transformed $r$ into the canonical radial field $\phi$. 
Substituting the definition \eqref{canonical field} into the Lagrangian \eqref{kinetic lagrangian 1} yields
\begin{equation}
    \label{kinetic lagrangian 2}
    \mathcal{L}=\frac{1}{2}R -\frac{1}{2}\left(\partial \phi \right)^2 -\frac{3\alpha}{4} \sinh^2{\left(\frac{2\phi}{\sqrt{6\alpha}} \right)} \left( \partial\theta \right)^2-c_2 \tanh^2{\left(\frac{\phi}{\sqrt{6\alpha}}\right)} - c_4 \tanh^4{\left(\frac{\phi}{\sqrt{6\alpha}}\right)} + \cdots\;.
\end{equation}
The angular field is non-canonical and when both $\phi$ and $\theta$ are light during inflation the full multi-field dynamics should be taken into account \cite{Achucarro:2017ing, Christodoulidis:2018qdw, Linde:2018hmx, Krajewski:2018moi, Iarygina:2018kee, Iarygina:2020dwe, Kallosh:2022ggf, Kallosh:2022vha}. Nevertheless, 
if the angular field $\theta$ is strongly stabilised during inflation \cite{Carrasco:2015uma} then it can be integrated out, leading to an effective single-field description of $\alpha$-attractors in terms of the radial field $\phi$, which plays the role of the inflaton. The coarse-grained version of Eq.~\eqref{kinetic lagrangian 2} is 
\begin{equation}
    \label{kinetic lagrangian 4}
    \mathcal{L}=\frac{1}{2}R -\frac{1}{2}\left(\partial \phi \right)^2-c_2 \tanh^2{\left(\frac{\phi}{\sqrt{6\alpha}}\right)} - c_4 \tanh^4{\left(\frac{\phi}{\sqrt{6\alpha}}\right)} +\cdots\;.
\end{equation}

We will focus on the case where
the lowest-order non-zero contribution dominates 
and the potential 
is given by Eq.~\eqref{single field potential}, where $p$ is even and we identify the coefficient of the leading term with $V_0$. 
This simple T-model potential \eqref{single field potential} then depends on the three parameters $\{V_0,\, \alpha, \, p\}$. The case $p=2$ corresponds to a simple mass term in the parent theory \cite{Kallosh:2013yoa,Kallosh:2015zsa}, see Eq.~\eqref{V of Z}. We will also consider the massless case where the leading order term corresponds to $p=4$.

The equations of motion for the background evolution are 
\begin{align}
\label{single field inflaton eom}
    \ddot \phi+3H\dot \phi+V_\phi&=0\;, \\
\label{phi dot with H dot}
    \dot H+ \frac{1}{2}\dot \phi^2 &=0 \;,
\end{align}
where $V_\phi\equiv{\mathrm{d}V(\phi)/\mathrm{d}\phi}$, and the system obeys the Friedmann constraint
\begin{equation}
\label{Friedmann single field}
    H^2=\frac{1}{3}\left[\frac{1}{2}\dot \phi^2 +V(\phi)\right] \;.
\end{equation}
The first slow-roll parameter is defined as 
\begin{equation}
    \epsilon_H \equiv -\frac{\dot H}{H^2}\;. 
\end{equation}
%
During inflation the field slowly rolls down the slope of its potential, which at $|\phi|\gg \sqrt{6\alpha}$ displays a plateau, and drives the background accelerated expansion, corresponding to $\epsilon_H<1$. 

When the inflaton leaves the plateau, it gains momentum, exits the slow-roll regime and starts oscillating. Inflation ends when the time-averaged $\epsilon_H$, see Eq.~\eqref{time averaged epsilonH}, becomes larger than unity.
For models with large $\alpha$ inflation ends when $\epsilon_H=1$ for the first time. Even if $\epsilon_H$ crosses unity multiple times during the subsequent oscillations, the time-averaged value of $\epsilon_H$ is greater than unity. For $\alpha\ll1$ the situation is slightly different. The time-averaged slow-roll parameter may remain smaller than unity, and inflation may be prolonged in the form of oscillating inflation \cite{Lin:2023ugk}. In practice the duration of the oscillatory inflation phase is short and 
we can neglect it in the following; see appendix \ref{app: inflaton oscillations} for more details. 

Once the inflaton oscillates about the minimum of its potential, 
the inflaton, and/or its decay products, must decay into Standard Model particles which rapidly thermalise. The process describing the energy transfer from the inflaton sector to ordinary matter goes by the name of reheating \cite{Kofman:1997yn}, which we discuss further in section \ref{sec: reheating intro}. Different $p$ values in Eq.~\eqref{single field potential} correspond to different shapes of the effective potential about its minimum\footnote{This approximation is not valid during the initial oscillatory inflation phase for models with small $\alpha$, see appendix \ref{app: inflaton oscillations}. Nevertheless, this phase is typically so short that, as mentioned above, we can neglect it and  approximate the potential as in Eq.~\eqref{potential small phi}.}, 
\begin{equation}
\label{potential small phi}
    V(\phi)\sim \left(\frac{\phi}{\sqrt{6\alpha}}\right)^p\;\;\; \text{for} \;|\phi|\ll \sqrt{6\alpha}  \;,
\end{equation}
which in turn affects the equation-of-state parameter, $w=P/\rho$, 
and hence the expansion of the universe during reheating. For T-models with potential \eqref{single field potential} the equation-of-state parameter during reheating is \cite{Turner:1983he}
\begin{equation}
\label{equation of state parameter}
    w=\frac{p-2}{p+2} \;.
\end{equation}
When the inflaton can be approximately described as an oscillating massive field ($p=2$), the background density redshifts like pressureless matter ($w=0$), and when the potential is approximated with a quartic function ($p=4$), the background behaves as during radiation domination ($w=1/3$).

\begin{figure}
    \centering
    \includegraphics[width = 0.6\textwidth]{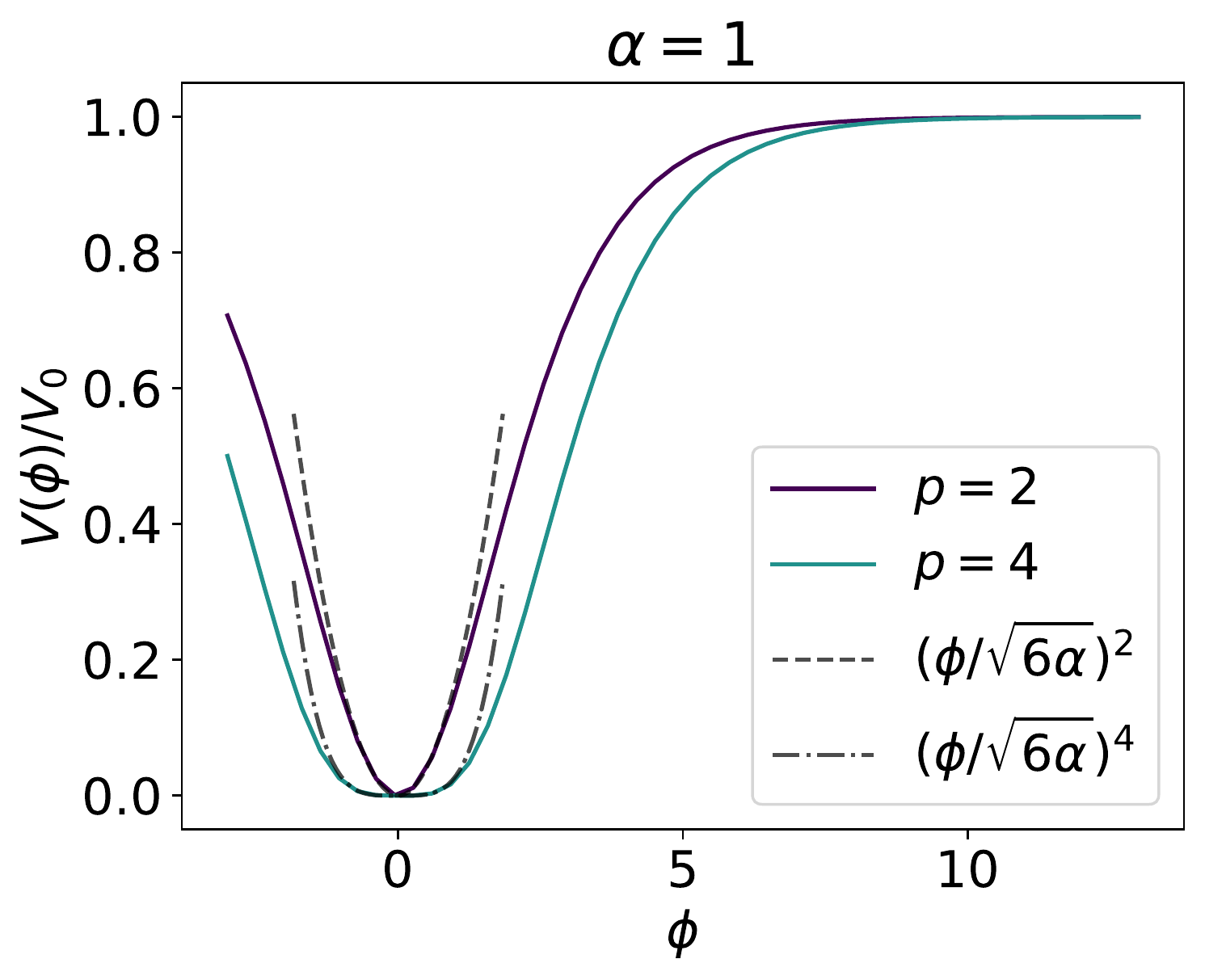}
    \caption{T-model potential \eqref{single field potential} with $\alpha=1$ and $p=\{2,\, 4\}$. At small inflaton values, we represent also Eq.~\eqref{potential small phi}, see the gray (dashed and dot-dashed) lines.}
    \label{fig:potential with different p}
\end{figure}
In this work we consider models with $p=2$ and $p=4$, whose potentials are represented in Figure \ref{fig:potential with different p} for $\alpha=1$. The potential displays a plateau for $\phi\gg \sqrt{6\alpha}$, while for $|\phi|\lesssim \sqrt{6\alpha}$ its shape is well described by Eq.~\eqref{potential small phi}. We start off by considering the case $p=2$ and work with models with $p=4$ in section \ref{sec:models with p=4}.

\subsection{Slow-roll inflation}
\label{sec:slow-roll approximation}
The large-scale dynamics of canonical, single-field models of inflation is well described by the slow-roll approximation, based on the assumption of an overdamped ($|\ddot \phi| \ll 3H|\dot \phi|$) and slowly rolling ($\dot \phi^2\ll V(\phi)$) inflaton field. For a comprehensive review of canonical, slow-roll, single-field models see \cite{Martin:2013tda}. 

The slow-roll approximation can be conveniently described by 
a hierarchical series of parameters \cite{Liddle:1994dx}, defined in terms of the inflaton potential, $V(\phi)$, or the Hubble rate during inflation, $H(\phi)$. We will refer to the two sets of slow-roll parameters as \textit{potential slow-roll parameters} (PSRP) and \textit{Hubble slow-roll parameters} (HSRP) respectively. The first three PSRS parameters are 
\begin{equation}
\label{PSRP}
    \epsilon_V\equiv \frac{1}{2}\left(\frac{V_\phi}{V} \right)^2 \;, \quad \eta_V\equiv \frac{V_{\phi\phi}}{V}\;, \quad {\xi_V}^2\equiv \frac{V_\phi V_{\phi\phi\phi}}{V^2} \;,
\end{equation}
while the first three HSRP\footnote{We express here the HSRP in terms of e-folding derivatives of the inflation field, instead of $H$, as this helps when dealing with numerical solutions.} are 
\begin{equation}
\label{HSRP phi N}
    \epsilon_H\equiv \frac{1}{2}\phi'^2\;, \quad \eta_H\equiv \frac{1}{2}\phi'^2 -\frac{\phi''}{\phi'}\;, \quad {\xi_H}^2\equiv \frac{1}{4}\phi'^4 -\frac{3}{2} \phi'\phi'' -\frac{\phi''^2}{\phi'^2} +\frac{\phi'''}{\phi'}\;,
\end{equation}
where $'\equiv \mathrm{d}/\mathrm{d}N$ and the elapsed number of e-folds is defined as $N\equiv \int \mathrm{d}t\, H(t)$. Formally, we will define the slow-roll regime by requiring all the parameters in the slow-roll hierarchy to be small.

Studying the inflaton dynamics under the slow-roll condition allows us to recover some analytical solutions. For example, we can reduce the inflaton equation of motion \eqref{single field inflaton eom} to
\begin{equation}
    \frac{\mathrm{d}\phi}{\mathrm{d}N}\simeq-\frac{V_\phi}{V} \;,
\end{equation}
where we transformed derivatives with respect to cosmic time into derivatives with respect to the number of e-folds. Integrating the equation above yields the number of e-folds elapsed between the two field values $\phi_\text{CMB}$ and $\phi_\text{end}$, 
\begin{equation}
\label{single field Delta N in terms of phi end and phi CMB}
    \Delta N_\text{CMB} \equiv N_\text{end}-N_\text{CMB} 
    \simeq \int_{\phi_\text{end}}^{\phi_\text{CMB}}  \mathrm{d}\phi\,\frac{V}{V_\phi}  \;.
\end{equation}
In the equation above, $\phi_\text{CMB}$ and $\phi_\text{end}$ represent the values of the inflaton field when the CMB scale crossed the horizon and at the end of inflation respectively, where we define the end of inflation to be the first occurrence of $\epsilon_H=1$ after the slow-roll evolution. By leaving $\phi_\text{end}$ explicit, integrating Eq.~\eqref{single field Delta N in terms of phi end and phi CMB} for the potential \eqref{single field potential} with $p=2$ yields
\begin{equation}
\label{phi CMB}
\sinh^2{\left(\frac{\phi_\text{CMB}}{\sqrt{6\alpha}}\right)} \simeq \frac{2 \Delta N_\text{CMB}}{3\alpha} + \frac{1}{2} \cosh{\left( \frac{2\phi_\text{end}}{\sqrt{6\alpha}}\right) }-\frac{1}{2} \;.
\end{equation}
The expression \eqref{phi CMB} can then be inverted to find $\phi_\text{CMB}$ given $\Delta N_\text{CMB}$. 

At leading order in the slow-roll approximation, the scalar power spectrum for primordial density perturbations is given in terms of the Hubble rate, $H$, and $\epsilon_H$ as \cite{ParticleDataGroup:2022pth}
\begin{equation}
\label{slow roll power spectrum}
    P_\zeta(k)=\frac{H^2}{8\pi^2 \epsilon_H} \Big|_{k=aH} \;,
\end{equation}
where this expression is evaluated for each scale at horizon crossing, $k=aH$. At leading order in slow-roll, $H^2\simeq V/3$ and $\epsilon_H\simeq \epsilon_V$, see Eq.~\eqref{exact epsilon_V}, therefore $P_\zeta(k)$ can be equivalently written in terms of the inflaton potential as
\begin{equation}
\label{slow roll power spectrum potential}
    P_\zeta(k)=\frac{V}{24\pi^2 \epsilon_V} \Big|_{k=aH} \;.
\end{equation}

For the purpose of comparing the primordial scalar power spectrum to current CMB data, it is useful to parametrise the scalar power spectrum on large scales with a simple power-law expression \cite{Planck:2018jri}, 
\begin{equation}
\label{P zeta power law}
    P_\zeta(k)=\mathcal{A}_s \left(\frac{k}{k_\text{CMB}} \right)^{n_s-1} \;,
\end{equation}
where $\mathcal{A}_s$ and $n_s-1$ are the amplitude and spectral tilt of $P_\zeta(k)$ respectively, defined at the scale $k_\text{CMB}=0.05\,\text{Mpc}^{-1}$.

Analogously, one can derive the slow-roll expression for the power spectrum describing the primordial GWs produced during inflation. In particular, the amplitude of the tensor power spectrum is usually discussed in terms of the tensor-to-scalar ratio, 
\begin{equation}
    \label{definition r}
    r\equiv \frac{\mathcal{A}_t}{\mathcal{A}_s} \;,
\end{equation}
where $\mathcal{A}_t=2H^2/\pi^2$ is the amplitude of the primordial tensor power spectrum at leading order in slow roll~\cite{ParticleDataGroup:2022pth}.

\subsection{Reheating}
\label{sec: reheating intro}
During reheating, the energy density decreases from its value at the end of inflation, $\rho_\text{end}$, to $\rho_\text{th}$, when the inflaton, and/or its decay products, have decayed into Standard Model particles and these are thermalised. The duration of the reheating phase and its effective equation-of-state have an impact on the CMB observables predicted by a specific inflationary model, see Eq.~\eqref{Nstar}, and the Bayesian evidence for a model changes depending on the assumptions made about the reheating phase \cite{Martin:2014nya}.  

The duration of reheating is measured in terms of e-foldings of expansion after the end of inflation, $\Delta \tilde N_\text{rh}\equiv N_\text{rh}-N_\text{end}$, where $N_\text{rh}$ marks the end of reheating and we use a tilde symbol to identify e-folds elapsed after the end of inflation. The quantity $\Delta \tilde N_\text{rh}$ depends on the effective equation-of-state parameter, $w$, and the value of $\rho_\text{th}$, and is given by\footnote{We note that the instantaneous equation-of-state parameter is not constant for the entire duration of reheating, e.g. changes from $-1$ at the end of inflation to $1/3$ at the beginning of radiation domination, while the effective equation-of-state parameter introduced in Eq.\eqref{N rh} is the mean value of it during the period of reheating (see e.g. \cite{Martin:2014nya}), and it is therefore a constant. For a work on the time evolution of the equation-of-state parameter see, e.g., \cite{Saha:2020bis}.
In practice, the dynamics during reheating can be quite complex, including e.g. non-perturbative phenomena such as parametric resonance, and it therefore constitutes a research topic on its own right (see e.g. \cite{Lozanov:2017hjm}). We work here within a simplified reheating scenario and leave the investigation of more complex processes for future work.}
\begin{equation}
    \label{N rh}
    \Delta \tilde N_\text{rh} \equiv \frac{1}{3(1+w)}\log{\left( \frac{\rho_\text{end}}{\rho_\text{th}}\right)} \;.
\end{equation}
For $p=2$ we have $w=0$, see Eq.~\eqref{equation of state parameter}, implying that reheating is well approximated by a matter-dominated expansion. In this case the thermalisation energy is given as 
\begin{equation}
{\rho_\text{th}}^{1/4}={\rho_\text{end}}^{1/4}\exp{\left(-3/4\,\Delta \tilde N_\text{rh}\right)}\;. 
\end{equation}
For works on constraining the reheating phase in $\alpha$-attractors and related models see e.g. \cite{Ueno:2016dim, Nozari:2017rta, Drewes:2017fmn, Mishra:2021wkm, Ellis:2021kad, Ling:2021zlj, Drewes:2022nhu, Drewes:2023bbs, Chakraborty:2023ocr}.

We place an upper bound on $\Delta \tilde N_\text{rh}$ by requiring that reheating is complete before the onset of Big-Bang nucleosynthesis, which implies\footnote{Recalling that nuclei form at energies $\sim 100\,\text{keV}$, the reheating temperature has to be larger than $\sim 1\,\text{MeV}$ to allow for successful Big Bang nucleosynthesis, while being likely much higher to allow for baryogenesis after inflation \cite{Baumann:2022mni}. We choose $\rho_\text{th}\geq \left(1\text{TeV}\right)^4$, since baryogenesis requires physics beyond the Standard Model \cite{Mukhanov:2005sc}, and experiments in particle accelerators have tested particle physics, and found that it is largely consistent with the Standard Model, up to energies $\sim 10\,\text{TeV}$ \cite{Collaboration:2022598}. We expect that extending the lower limit on $\rho_\text{th}$ to smaller values would not impact the conclusions of our analysis, since smaller values of $\rho_\text{th}$, i.e. longer reheating stages, would push $n_s$ to even smaller values (see Figures \ref{fig:ns r slow roll} and \ref{fig:Planck likelihood}), incompatible with CMB experiments.} $\rho_\text{th}\in \left[\left(1\text{TeV}\right)^4, \, \rho_\text{end}\right]$ \cite{Planck:2018jri}, where $\rho_\text{th}=\rho_\text{end}$ corresponds to the case of instant reheating. This leads to an upper bound on the duration of reheating, 
\begin{equation}
\label{N rh max}
    \Delta \tilde N_\text{rh} \leq \frac{1}{3}\log{\left( \frac{\rho_\text{end}}{\left(1\text{TeV}\right)^4}\right)} \equiv \Delta \tilde N_\text{rh,max}  \;.
\end{equation}
\begin{figure}
\centering
\includegraphics[width=.6\textwidth]{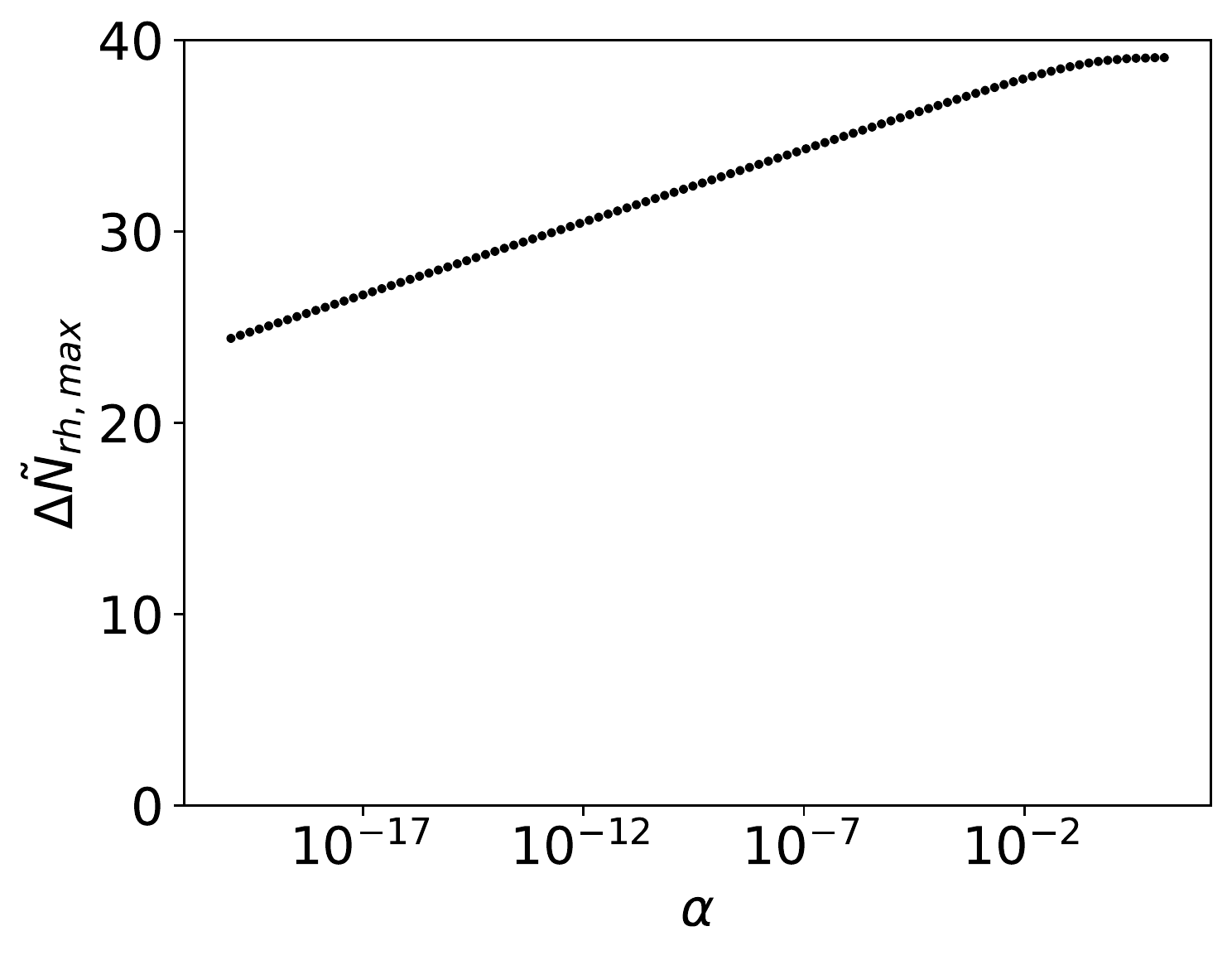}
\caption{Maximum duration of reheating for $\alpha$-attractor T-models with potential \eqref{single field potential}, $p=2$ and $10^{-20}\leq\alpha\leq15$. The points are produced numerically by using the definition of $\Delta \tilde N_\text{rh,max}$ in Eq.~\eqref{N rh max}.}
\label{fig:max duration of reheating}
\end{figure}

For T-models \eqref{single field potential} with $p=2$ and $10^{-20}\leq\alpha\leq15$, we numerically calculate the energy density at the end of inflation, $\rho_\text{end}$, and, by means of Eq.~\eqref{N rh max}, represent $\Delta \tilde N_\text{rh,max}$ in Figure \ref{fig:max duration of reheating}. We find that $\Delta \tilde N_\text{rh,max}$ depends logarithmically on $\alpha$, with smaller $\alpha$ corresponding to shorter allowed reheating stages. This is due to the fact that models with small $\alpha$ require smaller $H$ during inflation, i.e. smaller $r$, see Eq.~\eqref{r universal}, which implies that also $\rho_\text{end}$ and therefore $\Delta \tilde N_\text{rh,max}$, see Eq.~\eqref{N rh max}, will be smaller. 

\section[\texorpdfstring{$\bm{\alpha}$-dependence of the CMB observables}{alpha-dependence of the CMB observables}]{$\bm{\alpha}$-dependence of the CMB observables}
\label{sec:alpha scaling of the CMB observables}

The $\alpha$-attractor standard universal prediction for the tensor-to-scalar ratio, Eq.~\eqref{r universal}, clearly shows an explicit dependence on the parameter $\alpha$. We demonstrate in this section that the scalar spectral index, Eq.~\eqref{ns universal}, also implicitly depends on $\alpha$ through the quantity $\Delta N_\text{CMB}$. 

In section \ref{sec: analytic Delta N CMB}, we derive an analytic expression for $\Delta N_\text{CMB}$ and use it to explore the overall dependence on $\alpha$ of the standard universal predictions, Eqs.\eqref{ns universal}-\eqref{r universal}. In section \ref{sec:improved predictions} we employ this result to improve the extended and standard universal predictions, Eqs.\eqref{ns p dependence}-\eqref{r p dependence} and Eqs.\eqref{ns universal}-\eqref{r universal} respectively, and compare them with numerical results for $n_s$ and $r$ for T-models with $p=2$ and $10^{-20}\leq\alpha\leq15$. We will motivate this range for $\alpha$ in section \ref{sec:CMB bounds}. 

\subsection[\texorpdfstring{An analytic expression for ${\Delta N_\text{CMB}(\alpha,\, \Delta \tilde N_\text{rh})}$}{An analytic expression for Delta N CMB(alpha, Delta N rh)}]{An analytic expression for $\bm{\Delta N_\text{CMB}(\alpha,\, \Delta \tilde N_\text{rh})}$}
\label{sec: analytic Delta N CMB}

The number of e-folds elapsed between the horizon crossing of the CMB scale, $k_\text{CMB}=aH$, and the end of inflation \cite{Planck:2018jri}
\begin{equation}
\begin{split}
\label{Nstar}
    \Delta N_\text{CMB} 
    &\equiv N_\text{end}-N_\text{CMB} \\
    &\simeq 61.02+\frac{1}{4}\ln{\left(\frac{V_\text{CMB}^2}{ \rho_\text{end}} \right) }-\frac{1-3w}{4} \Delta \tilde N_\text{rh} \,,  
\end{split}
\end{equation}
is a key quantity for the phenomenology of an inflationary model. In this expression $V_\text{CMB}$ is the value of the potential when the comoving wavenumber $k_\text{CMB}$ crossed the horizon during inflation. We have already introduced the quantities $\rho_\text{end}$, $w$ and $\Delta \tilde N_\text{rh}$ in section \ref{sec: reheating intro}. We note that in \eqref{Nstar} we have set the number of effective bosonic degrees of freedom at the completion of reheating to $10^3$ as done in \cite{Planck:2018jri}\footnote{Since at energies above $\sim 10\,\text{TeV}$ we expect deviations from the physics of the Standard Model, e.g. to account for baryogenesis (see footnote 4), we accordingly choose here a number of effective bosonic degrees of freedom, $g_\text{th}$, beyond the Standard Model one, $g_\text{th,SM}=106.75$. We also note that $g_\text{th}$ contributes to $\Delta N_\text{CMB}$ through a factor $1/12 \ln{(g_\text{th})}$ \cite{Planck:2013jfk}, which is small and does not strongly depend on the particle content at the end of reheating, for example $1/12 \ln{(g_\text{th,SM})}\sim 0.39$ and $1/12 \ln{(10^3)}\sim 0.57$.}. The precise value of $\Delta N_\text{CMB}$ thus depends on the inflationary potential and the details of reheating \cite{Martin:2014nya}. For reheating stages with $-1/3<w<1/3$, the value of $\Delta N_\text{CMB}$ is maximised for instant reheating, $\rho_\text{th}=\rho_\text{end}$.  

In the following we make explicit the dependence of $\Delta N_\text{CMB}$ on the model parameter $\alpha$ for T-models \eqref{single field potential} with $p=2$. For convenience we first assume instant reheating (and reintroduce later an extended reheating phase, $\Delta \tilde N_\text{rh}\neq0$, see Eq.~\eqref{N CMB with rh}). In this case Eq.~\eqref{Nstar} can be written as 
\begin{equation}
\label{Nstar step1}
    \Delta N_\text{CMB,inst rh}\simeq 61.02 +\frac{1}{4}\ln{\left(\frac{V_\text{CMB}^2}{ \rho_\text{end}} \right) } \;,
\end{equation}
where the numerator in the second term is evaluated $\Delta N_\text{CMB,inst rh}$ e-folds before the end of inflation, i.e. at $\phi=\phi_\text{CMB}$, and the denominator is evaluated at the end of inflation, i.e. at $\phi=\phi_\text{end}$. 
The energy density at the end of inflation is 
\begin{equation}
    \rho_\text{end}=\frac{1}{2}{\dot{\phi}_\text{end}}^2 +V(\phi_\text{end}) = \frac{3}{2}V(\phi_\text{end}) \;,
\end{equation}
where Eq.~\eqref{Friedmann single field} and the condition ${\dot \phi_\text{end}}^2/2= {H_\text{end}}^2$ at the end of inflation have been used. Substituting this into Eq.~\eqref{Nstar step1} yields
\begin{equation}
\label{Nstar step2}
    \Delta N_\text{CMB,inst rh}\simeq 60.92 +\frac{1}{4}\ln{\left(\frac{V_\text{CMB}^2}{ V_\text{end}} \right) } \;.
\end{equation}
This expression depends on $\phi_\text{CMB}$, $\phi_\text{end}$ and on the potential normalisation $V_0$. The value of $\phi_\text{CMB}$ can be calculated for a given $\phi_\text{end}$ in the slow-roll approximation, see Eq.~\eqref{phi CMB}, while the potential normalisation, $V_0$, is fixed at CMB scales by measurements of the amplitude of the scalar power spectrum, $\mathcal{A}_s=2.1\times 10^{-9}$ \cite{Planck:2018jri}, see Eq.~\eqref{P zeta power law}. Using the T-model potential \eqref{single field potential} with $p=2$ in Eq.~\eqref{slow roll power spectrum potential} and the aforementioned normalisation yields 
\begin{equation}
\label{V0 normalisation single field}
    V_0\simeq 6\times 10^{-6} \alpha {\left(-3\alpha +3\alpha \cosh{\left( \frac{2\phi_\text{end}}{\sqrt{6\alpha}}\right) } +4 \Delta N_\text{CMB,inst rh}  \right)^{-2}} \;.
\end{equation}
Substituting \eqref{single field potential}, \eqref{phi CMB} and \eqref{V0 normalisation single field} into Eq.~\eqref{Nstar step2} yields 
\begin{align}
\label{Nstar stage3}
    \Delta N_\text{CMB,inst rh}&= 57.56 +\frac{1}{4}\ln{\left[4\alpha \tanh^{-2}{\left(\frac{\phi_\text{end}}{\sqrt{6\alpha}} \right)} {\left(3\alpha +3\alpha \cosh{\left( \frac{2\phi_\text{end}}{\sqrt{6\alpha}}\right) } +4 \Delta N_\text{CMB,inst rh}\right)^{-2}} \right]} \;.
\end{align}
The value of the inflaton at the end of inflation, $\phi_\text{end}$, can be calculated by noting that at the end of inflation \cite{Ellis:2015pla}
\begin{equation}
    \label{epsilon_v end of inflation}
    \epsilon_V\simeq \left(1+\sqrt{1-\eta_V/2} \right)^2 \;.
\end{equation} 
By calculating $\epsilon_V$ and $\eta_V$ for the potential \eqref{single field potential} with $p=2$, one can solve Eq.~\eqref{epsilon_v end of inflation} to obtain 
\begin{equation}
    \label{phi end}
    \phi_\text{end}(\alpha)= \sqrt{\frac{3\alpha}{2}}\,\coth^{-1}{\left[\frac{2 \left(10 +16 \alpha +17\sqrt{\alpha (5+4\alpha)} \right)}{25 \sqrt{3}}  \sqrt{\frac{5+68\alpha -16 \sqrt{\alpha (5+4\alpha)}}{(1-12\alpha)^2}}\right]} \;.
\end{equation}

Equipped with an analytic expression for $\phi_\text{end}$, we can now solve Eq.~\eqref{Nstar stage3} and obtain an analytic expression for $\Delta N_\text{CMB,inst rh}$. By defining 
\begin{align}
\label{A1}
    A_1 &\equiv 57.56 + \frac{1}{4} \ln{\left[4\alpha \tanh^{-2}{\left(\frac{\phi_\text{end}}{\sqrt{6\alpha}} \right)} \right]} -\ln2\;,\\
\label{A2}    
    A_2 &\equiv \frac{3\alpha}{4} +\frac{3\alpha}{4} \cosh{\left( \frac{2\phi_\text{end}}{\sqrt{6\alpha}}\right) }\;,
\end{align}
Eq.~\eqref{Nstar stage3} can be recast into the form
\begin{equation}
    \label{Lambert 2}
    \Delta N_\text{CMB,inst rh} = {A}_1 -\frac{1}{2} \ln{\left( \Delta N_\text{CMB,inst rh} +{A}_2 \right)} \;,
\end{equation}
which admits a solution in terms of the Lambert function\footnote{Since $2\left(\Delta N_\text{CMB} +{A}_2 \right)$ and $2 \text{e}^{2\left({A}_1+{A}_2 \right)}$ are real and $2 \text{e}^{2\left({A}_1+{A}_2 \right)}>0$, the solution Eq.~\eqref{Lambert 3} is unique.} $W_0(x)$ (see also \cite{Ellis:2015pla, Ellis:2021kad}), 
\begin{equation}
\label{Lambert 3}
    2\left(\Delta N_\text{CMB} +{A}_2 \right) = W_0\left(2 \text{e}^{2\left({A}_1+{A}_2 \right)} \right) \;.
\end{equation}
By expanding $W_0(x)\simeq \ln{x}-\ln{\left(\ln{x} \right)}$, Eq.~\eqref{Lambert 3} yields
\begin{equation}
\label{N CMB}
    \Delta N_\text{CMB,inst rh}(\alpha)= A_1 +\frac{1}{2} \ln{2} -\frac{1}{2} \ln{\left(2A_1 +2A_2 +\log{2} \right)} \;,
\end{equation}
where $A_1$ and $A_2$ are defined in Eqs.~\eqref{A1} and \eqref{A2}. 
\begin{figure}
\centering
\includegraphics[width=.6\textwidth]{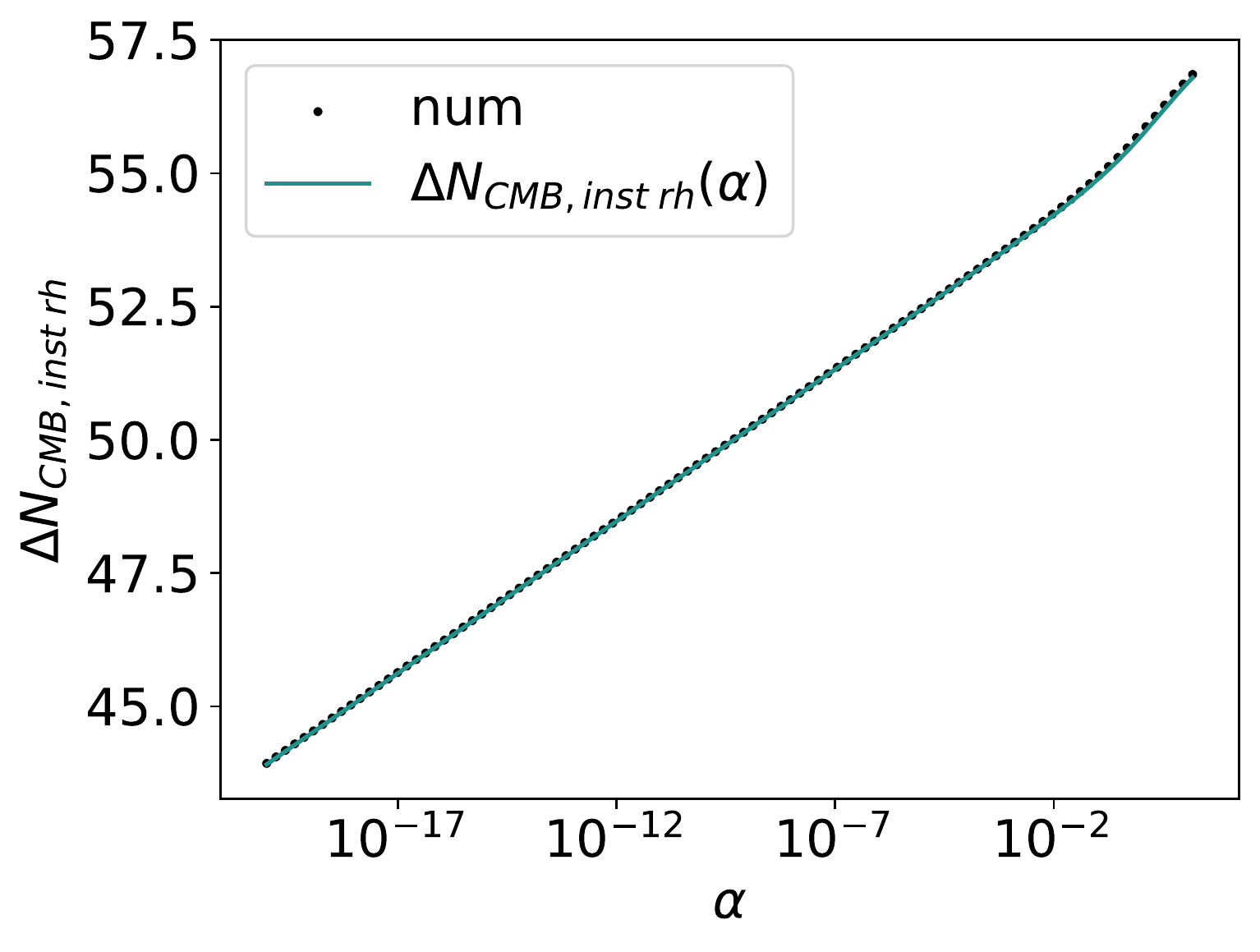}
\caption{Comparison between the numerical results (black dots) for $\Delta N_\text{CMB,inst rh}$ and the analytic expression $\Delta N_\text{CMB,inst rh}(\alpha)$ in Eq.~\eqref{N CMB} (green line) for $\alpha$-attractor T-models with potential \eqref{single field potential}, $p=2$ and $10^{-20}\leq \alpha\leq 15$. The numerical results are obtained under the assumption of instant reheating.}
\label{fig:N CMB}
\end{figure}
In Figure \ref{fig:N CMB} we compare the numerical results for $\Delta N_\text{CMB,inst rh}$, obtained under the assumption of instant reheating by iteratively solving \eqref{Nstar} with $V_0$ values compatible with CMB measurements, with the analytic expression \eqref{N CMB}. The analytic results provide a very good approximation for the numerical values. 

Figure \ref{fig:N CMB} shows that $\Delta N_\text{CMB,inst rh}$ displays a logarithmic dependence on the parameter $\alpha$. Even if the dependence is mild, this does impact on large-scale observables. Before turning to calculate the numerical predictions for the scalar spectral tilt and tensor-to-scalar ratio, which is the subject of section~\ref{sec:improved predictions}, let us sketch here the qualitative effect that $\Delta N_\text{CMB,inst rh}(\alpha)$ has on $n_s$ and $r$ by using the $\alpha$-attractor standard universal predictions, Eqs.~\eqref{ns universal} and \eqref{r universal}. 
\begin{figure}
    \centering
    \includegraphics[width=1\textwidth]{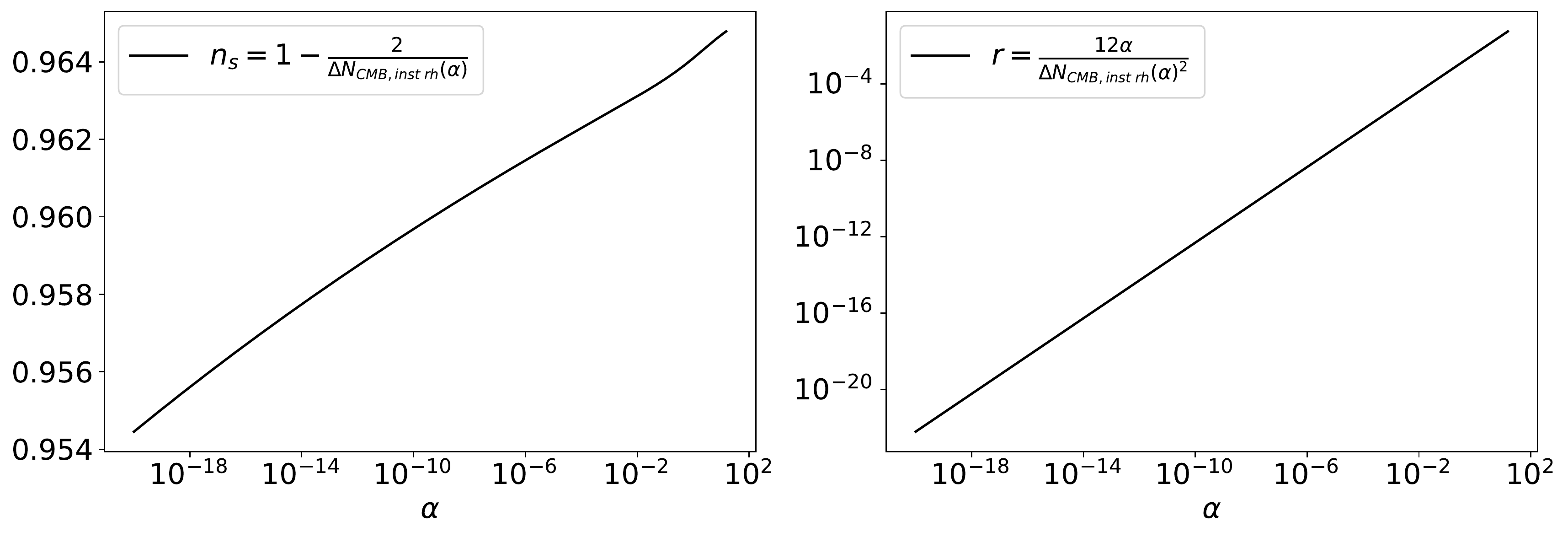}
    \caption{The $\alpha$-attractors standard universal predictions for the spectral tilt (left panel) and tensor-to-scalar ratio (right panel), see Eqs.~\eqref{ns universal} and \eqref{r universal} respectively, where the dependence on the parameter $\alpha$ is made explicit by means of Eq.~\eqref{N CMB}, derived for T-models with potential \eqref{single field potential}, $p=2$ and instant reheating.}
    \label{fig:ns and r universal predictions}
\end{figure}
In Figure \ref{fig:ns and r universal predictions} we show $n_s$ (left panel) and $r$ (right panel) against $\alpha$, obtained by substituting the expression in Eq.~\eqref{N CMB} for $\Delta N_\text{CMB,inst rh}(\alpha)$ into Eqs.~\eqref{ns universal} and \eqref{r universal}. In the left panel, $n_s$ inherits a logarithmic dependence on $\alpha$, with larger values corresponding to larger $\alpha$. On the other hand, $r$ displays a linear dependence on $\alpha$ (note the axes used in the right panel). This is due to the explicit $\alpha$-dependence in the numerator of \eqref{r universal}, which is stronger than the logarithmic dependence on $\alpha$ appearing in the denominator through $\Delta N_\text{CMB,inst rh}(\alpha)$. 

While it is well known that experimental upper bounds on $r$ can be used to limit $\alpha$ from above (because of the linear dependence of the tensor-to-scalar ratio on $\alpha$), we will show in section \ref{sec:CMB bounds} that the implicit dependence of $\Delta N_\text{CMB}$ on $\alpha$ can lead to additional constraints on $\alpha$ from measurements of the spectral tilt, $n_s$.

Let us also provide here an analytic expression for $\Delta N_\text{CMB}$ which allows for an extended reheating stage. By setting $w=0$ in Eq.~\eqref{Nstar}, it is possible to isolate\footnote{In the case of extended reheating, $\rho_\text{th}\neq\rho_\text{end}$, applying to Eq.~\eqref{Nstar} the same procedure used to get Eq.~\eqref{N CMB} yields
\begin{equation}
     \Delta N_\text{CMB}(\alpha, \, \Delta \tilde N_\text{rh} ) = \Delta N_\text{CMB,inst rh}(\alpha)-\frac{1}{4} \Delta \tilde N_\text{rh} -\frac{1}{2} \log{\left(1-\frac{\Delta \tilde N_\text{rh}}{4A_1+4A_2+2\log2} \right)}\;,
\end{equation}
where $A_1$ and $A_2$ are defined in Eqs.~\eqref{A1} and \eqref{A2} respectively. Given that the third term on the right-hand side is roughly two orders of magnitude smaller than the second one in the range of $\alpha$ considered, we can safely approximate $\Delta N_\text{CMB}(\alpha, \, \Delta \tilde N_\text{rh} )$ with the simpler expression in Eq.~\eqref{N CMB with rh}.} the reheating contribution to the value of $\Delta N_\text{CMB}$, 
\begin{equation}
    \label{N CMB with rh}
    \Delta N_\text{CMB}(\alpha, \, \Delta \tilde N_\text{rh} )\simeq \Delta N_\text{CMB,inst rh}(\alpha)-\frac{1}{4} \Delta \tilde N_\text{rh} \;,
\end{equation}
where we identify $\Delta N_\text{CMB,inst rh}(\alpha)$ with the analytic expression \eqref{N CMB}. Eq.~\eqref{N CMB with rh} shows that the presence of a finite reheating phase with $w<1/3$ decreases the value of $\Delta N_\text{CMB}$, with consequences for the large-scale observables. In particular, $\Delta \tilde N_\text{rh}\neq0$ always decreases the prediction for $n_s$ with respect to the case $\Delta \tilde N_\text{rh}=0$, while the change produced in $r$ is not appreciable, moving the observables to the left in the $(n_s, \,r)$ plane.

By using Eq.~\eqref{N rh max} and ~\eqref{N CMB with rh}, one can calculate the allowed range of $\Delta N_\text{CMB}$ predicted for these models: $37.8\leq \Delta N_\text{CMB}\leq 56.8$.

\subsection{Improved predictions for the large-scale observables}
\label{sec:improved predictions} 

By using Eq.~\eqref{N CMB with rh}, we can correctly include the dependence on $\alpha$ and $\Delta \tilde N_\text{rh}$ inside of the extended and standard universal predictions. We obtain for the extended predictions~\eqref{ns p dependence}-\eqref{r p dependence},
\begin{align}
\label{ns new p dependence}
    n_s(\alpha, \, \Delta \tilde N_\text{rh}) &\approx 1-\frac{2\Delta N_\text{CMB}(\alpha, \, \Delta \tilde N_\text{rh})+\frac{1}{p}\sqrt{3\alpha(3\alpha+p^2)}+\frac{3\alpha}{2}}{\Delta N_\text{CMB}(\alpha, \, \Delta \tilde N_\text{rh})^2 +\frac{\Delta N_\text{CMB}(\alpha, \, \Delta \tilde N_\text{rh})}{p}\sqrt{3\alpha(3\alpha+p^2)}+\frac{3\alpha}{4}} \;, \\
\label{r new p dependence}
    r(\alpha, \, \Delta \tilde N_\text{rh}) & \approx \frac{12\alpha}{\Delta N_\text{CMB}(\alpha, \, \Delta \tilde N_\text{rh})^2 +\frac{\Delta N_\text{CMB}(\alpha, \, \Delta \tilde N_\text{rh})}{p}\sqrt{3\alpha(3\alpha+p^2)}+\frac{3\alpha}{4}}  \;, 
\end{align}
where for the models under consideration $p=2$, and for the universal predictions~\eqref{ns universal}-\eqref{r universal}
\begin{align}
    \label{ns new universal}
    n_s(\alpha, \, \Delta \tilde N_\text{rh})& \approx 1-\frac{2}{\Delta N_\text{CMB}(\alpha, \, \Delta \tilde N_\text{rh} )} \;, \\
    \label{r new universal}
    r(\alpha, \, \Delta \tilde N_\text{rh})& \approx \frac{12 \alpha}{\Delta N_\text{CMB}(\alpha, \, \Delta \tilde N_\text{rh} )^2} \;.
\end{align}
\begin{figure}
    \centering
    \includegraphics[width=.7\textwidth]{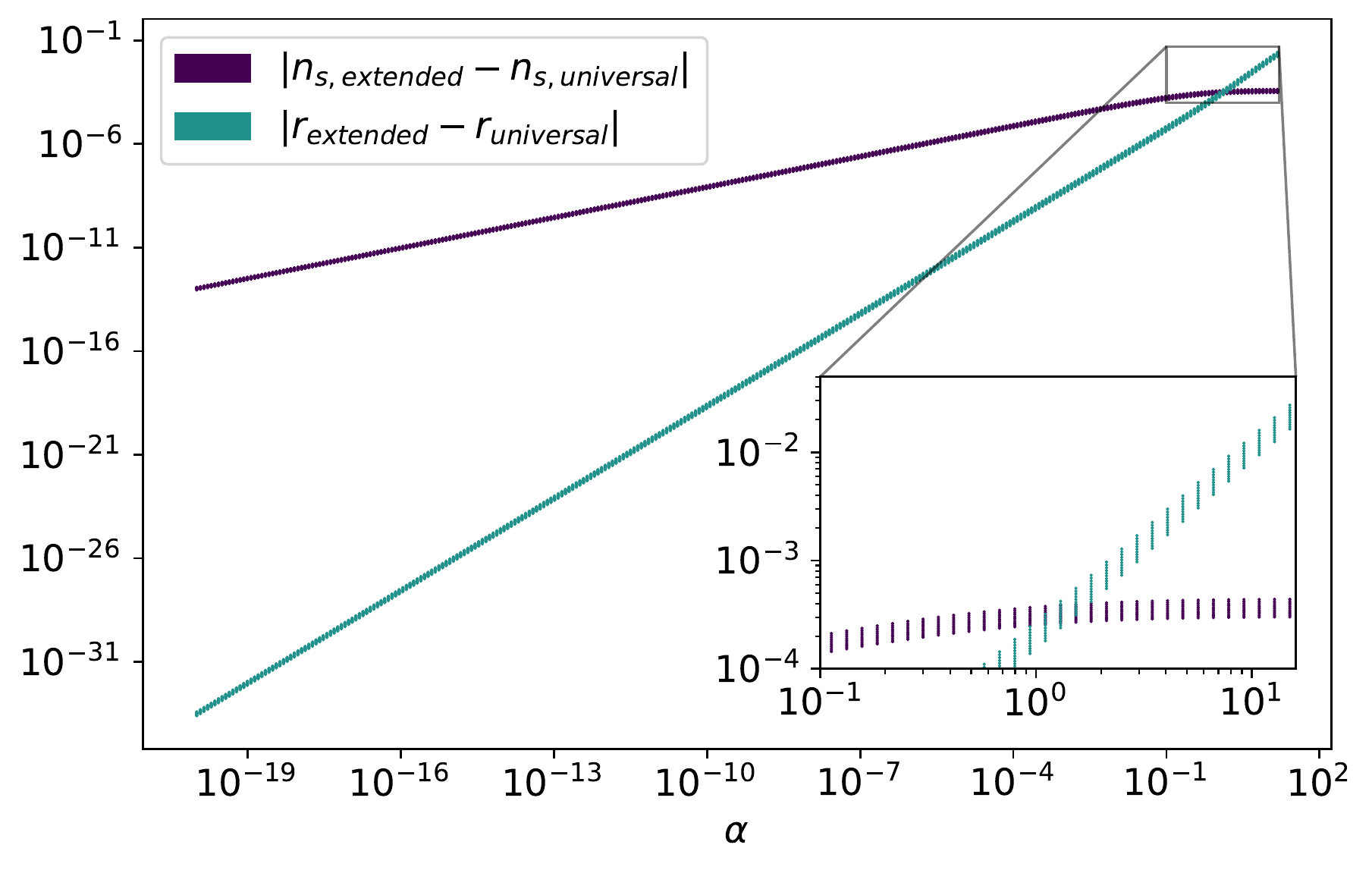}
    \caption{Difference between the $n_s$ and $r$ values obtained using the improved extended predictions, Eqs.\eqref{ns new p dependence}-\eqref{r new p dependence}, and the improved universal predictions, Eqs.\eqref{ns new universal}-\eqref{r new universal}. We consider T-models \eqref{single field potential} with $p=2$ and $10^{-20}\leq \alpha \leq 15$. For each $\alpha$, we consider 10 values of the reheating duration in the range $0\leq \Delta \tilde N_\text{rh}\leq \Delta \tilde N_\text{rh,max}(\alpha)$ (see Figure \ref{fig:max duration of reheating}).}
    \label{fig:difference between extended and universal predictions}
\end{figure}

In the following we will refer to these as the \textit{improved} extended and \textit{improved} universal predictions respectively. In Figure \ref{fig:difference between extended and universal predictions} we show the difference in the $n_s$ and $r$ values between the two types of prediction, where we use $p=2$ in Eqs.\eqref{ns new p dependence}-\eqref{r new p dependence}. When $\alpha$ is small, the results are practically indistinguishable. This is expected considering that the universal predictions are obtained from the extended expressions in the limit $\Delta N_\text{CMB}\gg \alpha, p$ (and at leading order in ${\Delta N_\text{CMB}}^{-1}$). For large $\alpha$, the explicit $\alpha$-dependent corrections in Eqs.\eqref{ns new p dependence}-\eqref{r new p dependence} become non-negligible and result in larger differences. In particular, for $\alpha=15$ we get\footnote{We quote here the maximum of the 10 values corresponding to different reheating durations.} $|n_{s,\text{extended}}-n_{s,\text{universal}}|\leq4.38\times 10^{-4}$ and $|r_{\text{extended}}-r_{\text{universal}}|\leq 2.7\times 10^{-2}$. 

\begin{figure}
    \centering
    \includegraphics[width = \textwidth]{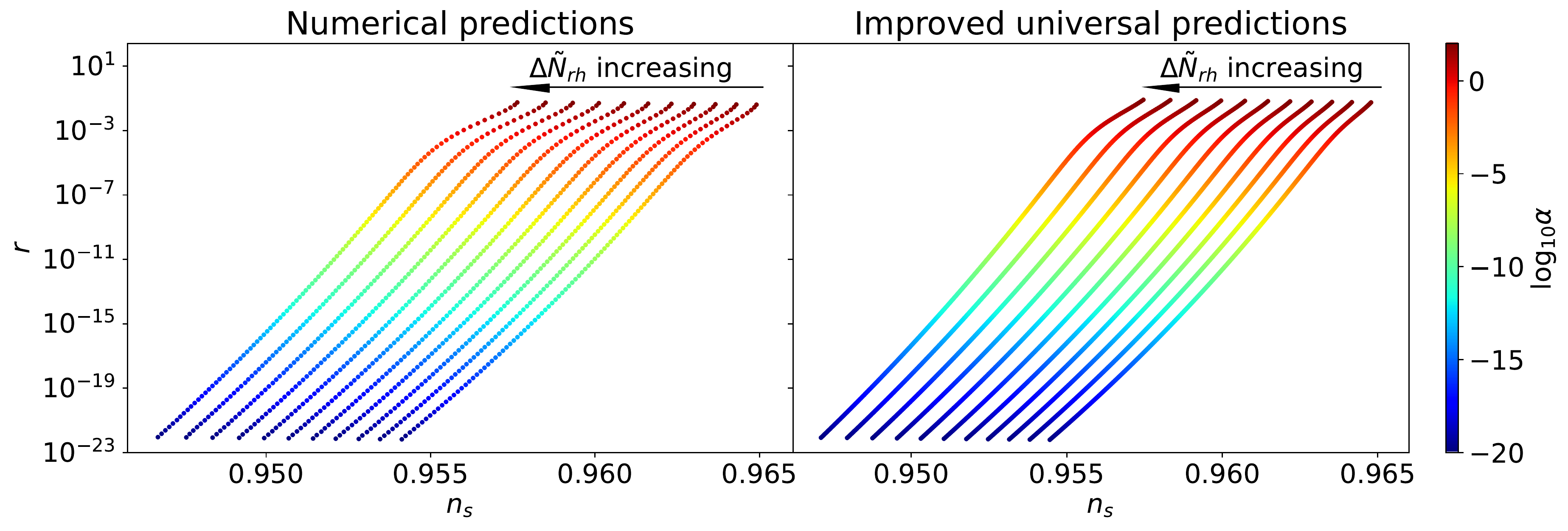}
    \caption{\textit{Left panel:} Numerical predictions for the large-scale observables in the $(n_s,\,r)$ plane for T-models \eqref{single field potential} with $p=2$, different values of $\alpha$, see the colored bar legend, and duration of reheating. Each of the tilted lines of points corresponds to a specific $\Delta \tilde N_\text{rh}$, with $\Delta \tilde N_\text{rh}$ increasing from $0$ to $\Delta \tilde N_\text{rh,max}(\alpha)$ from right to left. \textit{Right panel:} Improved universal predictions, see Eqs.\eqref{ns new universal}-\eqref{r new universal}, for the same models considered in the left panel.}
    \label{fig:ns r slow roll}
\end{figure}
\begin{figure}
    \centering
    \includegraphics[width = \textwidth]{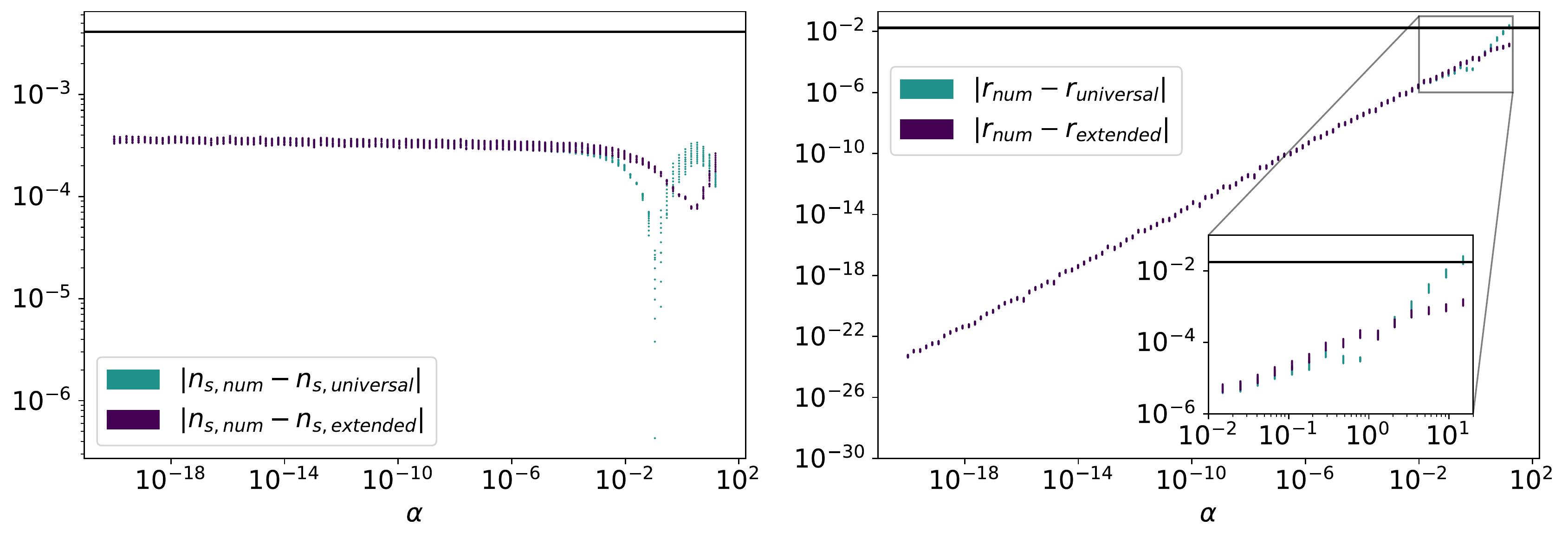}
    \caption{For each model whose predictions are represented in Figure \ref{fig:ns r slow roll}, we show here the absolute value of the difference between the numerical results and the improved extended (universal) predictions, see Eqs.\eqref{ns new p dependence}-\eqref{r new p dependence} (see Eqs.\eqref{ns new universal}-\eqref{r new universal}). For each $\alpha$, different points corresponds to a different reheating duration, in the range $0\leq \Delta \tilde N_\text{rh}\leq \Delta \tilde N_\text{rh,max}(\alpha)$. In the left (right) panel, the horizontal line represents $\sigma(n_s)$ ($\sigma(r)$), see Eq.\eqref{ns planck+BK18} (Eq.\eqref{r bound planck+BK18}).}
    \label{fig:check improved uni predictions}
\end{figure}

In Figure \ref{fig:ns r slow roll} we compare the improved universal\footnote{We quantitatively compare both the types of improved predictions with the numerical results in Figure \ref{fig:check improved uni predictions}.} predictions with numerical results for $n_s$ and $r$. In each case, we evaluate $n_s$ and $r$ for T-models \eqref{single field potential} with $p=2$, $10^{-20}\leq \alpha\leq15$ and, for each $\alpha$, 11 values of the reheating duration in the range $0\leq \Delta \tilde N_\text{rh}\leq \Delta \tilde N_\text{rh,max}(\alpha)$ (see Figure \ref{fig:max duration of reheating}). As discussed in appendix \ref{app:slow-roll observables}, we numerically evaluate $n_s$ at second-order in slow-roll, see Eq.~\eqref{ns second order HSRP}, and $r$ at first order, see Eq.~\eqref{r first order HSRP}. Henceforth we refer to these as the numerical predictions for $n_s$ and $r$. The similarity between the two panels is striking.

The results of Figure \ref{fig:ns r slow roll} show that $n_s$ has a logarithmic dependence on $\alpha$, which causes the predictions to be tilted in the $(n_s,\,r)$ plane. For a fixed duration of reheating, i.e. for each one of the tilted lines of points, we recognise the $\alpha$ dependence in the values of $r$. For fixed $\alpha$, an extended reheating stage moves the observables towards the left of the plane, due to the impact that $\Delta \tilde N_\text{rh}$ has on $n_s$, see Eq.~\eqref{ns new universal}, while the value of $r$ is not significantly changed, i.e. values of the tensor-to-scalar ratio calculated for fixed $\alpha$ and different $\Delta \tilde N_\text{rh}$ are all of the same order of magnitude. 
At this level of accuracy
the numerical predictions, as well as the improved analytic predictions, should not be approximated by vertical bands in the $(n_s,\,r)$ plane
due to the $\alpha$-dependence of $n_s$. 

When comparing the inflationary predictions with CMB data, analytical expressions can be used to describe the numerical results if the difference between the two is much smaller than $\sigma(n_s)$ and $\sigma(r)$ for the CMB survey considered. In this work we focus on current \textit{Planck} \cite{Planck:2018nkj, Planck:2019nip} and BICEP/Keck \cite{ BICEPKeck:2021gln} data, see the resulting $n_s$ measurement \eqref{ns planck+BK18} and $r$ upper bound \eqref{r bound planck+BK18}. We find that the difference between $n_{s,\text{num}}$ and the improved analytical predictions \eqref{ns new p dependence} and \eqref{ns new universal} is always at least one order of magnitude smaller than $\sigma(n_s)$ in \eqref{ns planck+BK18}. The same holds when comparing the numerical $r$ with the extended prediction \eqref{r new p dependence}. On the other hand, for large $\alpha$ the difference with respect to the improved universal prediction, $|r_\text{num}-r_\text{uni}(\alpha, \, \Delta \tilde N_\text{rh})|$, is comparable with $\sigma(r)$. 
These results are represented in Figure \ref{fig:check improved uni predictions}, and show that the improved extended predictions \eqref{ns new p dependence}-\eqref{r new p dependence} can be safely used when comparing the models with \textit{Planck} and BICEP/Keck 2018 data. On the other hand, due to the difference in $r$ for large $\alpha$ values, the improved universal predictions \eqref{ns new universal}-\eqref{r new universal} should not be used. We will take these considerations into account in section \ref{sec:CMB bounds}.

\section[\texorpdfstring{CMB constraints on $\bm{\alpha}$ and the duration of reheating}{CMB constraints on alpha and the duration of reheating}]{CMB constraints on $\bm{\alpha}$ and the duration of reheating}
\label{sec:CMB bounds}

In this section we confront the $\alpha$-attractor model predictions for the large-scale observables $n_s$ and $r$ with current CMB data. \textit{Planck} 2018 data \cite{Planck:2018nkj, Planck:2019nip} used together with BICEP/Keck 2018 data \cite{ BICEPKeck:2021gln} give the following contraints on $n_s$ and $r$ at $k_\text{CMB}=0.05\,\text{Mpc}^{-1}$ in the $\Lambda\text{CDM}+r$ cosmology \cite{Paoletti:2022anb}
\begin{gather}
    \label{ns planck+BK18}
    n_s=0.9653 \pm 0.0041 \quad (68\% \, \text{C.L.}) \;, \\
    \label{r bound planck+BK18}
    r<0.035 \quad (95\% \, \text{C.L.}) \;.
\end{gather}
The two-dimensional likelihood\footnote{The authors are very grateful to Daniela Paoletti for providing the original MCMC chains produced using the \texttt{CosmoMC} sampler \cite{Lewis:2002ah}, which \textit{Planck} and BICEP/Keck 2018 constraints on the standard $\Lambda$CDM + $r$ model in \cite{Paoletti:2022anb} were based on. From those MCMC chains, we produced a numerical, marginalized (over other cosmological parameters and \textit{Planck} plus BICEP/Keck nuisance parameters) likelihood for $n_s$ and $r$ using the \texttt{GetDist} package \cite{Lewis:2019xzd}.} is shown in Figure \ref{fig:Planck likelihood}.
We will examine model parameter constraints coming from the improved extended predictions, Eqs.\eqref{ns new p dependence}-\eqref{r new p dependence}, and compare these with parameter constraints derived using the standard universal predictions, Eqs.\eqref{ns universal}-\eqref{r universal}, in sections \ref{sec:improved extended predictions bayesian analysis} and \ref{sec: naive universal predictions bayesian analysis} respectively.

Our task is straightforwardly phrased in the language of Bayesian statistics. We have some prior (theoretical) knowledge about a set of fundamental parameters $X$, encoded in the prior probability $\text{pr}(X)$. When using the improved extended predictions, Eqs.\eqref{ns new p dependence}-\eqref{r new p dependence}, the fundamental parameters we want to constrain are $\alpha$ and $\Delta \tilde N_\text{rh}$, while if we use the standard universal predictions, Eqs.~\eqref{ns universal} and~\eqref{r universal}, the parameters are $\alpha$ and $\Delta N_\text{CMB}$.
If the underlying theory is characterised by additional symmetries \cite{Ferrara:2016fwe, Kallosh:2017ced}, $\alpha$ takes specific discrete values, $3\alpha=\{1,\,2,\, 3,\, 4,\, 5,\, 6,\, 7\}$, which may constitute a target for future CMB experiments \cite{Kallosh:2019eeu, Kallosh:2019hzo}. On the other hand, in principle $\alpha$ may take arbitrary values in the context of $\alpha$-attractors formulations in supergravity models and the magnitude of $\alpha$ is unbounded from below. Possible effects of small $\alpha$ have been investigated in \cite{Christodoulidis:2018qdw, Iacconi:2021ltm, Kallosh:2022vha}. In the absence of a theoretical prior on the order of magnitude of $\alpha$, we should take a uniform prior for $\log_{10}(\alpha)$. Alternatively if we expect $\alpha$ to take values of order one (in Planck units
) then we should consider a uniform prior which is linear in $\alpha$.
Note that by comparing the $95\%$ C.L. lower limit\footnote{We note that in a previous study \cite{German:2020cbw}, a lower bound on $r$, and consequently a lower bound on $\alpha$ (expressed in terms of $\lambda\equiv 1/\sqrt{6\alpha}$), was derived by requiring that $n_s$ is larger than the $68\%$ C.L. experimental lower bound. These are respectively $r>\mathcal{O}(10^{-10})$ and $\alpha>\mathcal{O}(10^{-8})$. The approach of \cite{German:2020cbw} differs from ours, as we choose the theoretical prior range to go beyond the $95\%$ C.L. experimental bound, and then derive observational bounds on the inflationary parameters $\alpha$ and $\Delta \tilde N_\text{rh}$ by performing a full Bayesian analysis.} on $n_s$, Eq.~\eqref{ns planck+BK18}, with Eq.~\eqref{ns new p dependence} for $\Delta \tilde N_\text{rh}=0,$\footnote{The most conservative lower limit on $\alpha$ is obtained for instant reheating scenarios since $\Delta \tilde N_\text{rh}\neq0$ moves $n_s$ to lower values thereby making the lower limit more stringent.} we obtain an approximate bound on $\alpha \gtrsim 5.7\times 10^{-16}$, while the upper bound on $r$, \eqref{r bound planck+BK18}, and Eq.~\eqref{r new p dependence} yield $\alpha\lesssim 12.7$. With these estimates in mind, we choose the prior range $\alpha\in[10^{-20},\,15]$. We expect this to extend beyond the observationally allowed range, and gives the range of values used previously in section \ref{sec:alpha scaling of the CMB observables}.

The prior knowledge of the model parameters, $\text{pr}(X)$, will be confronted with new  data $\mathcal{D}$, which we characterise with a likelihood function $\mathcal{L}(\mathcal{D}|X)$. We employ the \textit{Planck} and BICEP/Keck 2018 likelihood in the $(n_s,\,r)$ plane, i.e. marginalised over all other parameters. 
%
The posterior distribution for $X$ is then given by
\begin{equation}
\label{posterior}
    \mathcal{P}(X|\mathcal{D}) = \frac{\mathcal{L}(\mathcal{D}|X)\,\text{pr}(X)}{P(\mathcal{D})} \;, 
\end{equation}
where $P(\mathcal{D})=\int \mathrm{d}X\, \mathcal{L}(\mathcal{D}|X)\,\text{pr}(X)$ is usually known as the \textit{evidence}. We utilise Markov chain Monte Carlo (MCMC) methods to sample the posterior distribution $\mathcal{P}(X|\mathcal{D})$, employing the sampling algorithm implemented in the code \texttt{emcee} \cite{Foreman-Mackey:2012any}. We analyse the MCMC chains using the \texttt{GetDist} package \cite{Lewis:2019xzd}.

\subsection{Improved extended predictions}
\label{sec:improved extended predictions bayesian analysis}

In this subsection we will use the improved extended predictions, Eqs.~\eqref{ns new p dependence} and~\eqref{r new p dependence}, where the theoretical parameters are $\alpha$ and $\Delta \tilde N_\text{rh}$. 
Therefore, in addition to choosing a logarithmic or linear prior for $\alpha$, in each case we take a uniform prior for the duration of reheating\footnote{Previous studies have imposed additional constraints on the duration of reheating from the overproduction of  supersymmetric dark matter particles~\cite{Ellis:2021kad}, but we choose to remain agnostic a priori about the duration of reheating and allow the observational data to constrain this.}
in the range $0\leq \Delta \tilde N_\text{rh}\leq \Delta \tilde N_\text{rh, max}(\alpha)$, see Figure \ref{fig:max duration of reheating}.
Thus we identify $X=\{\alpha^*, \,\Delta \tilde N_\text{rh}\}$, where we have introduced the new variable $\alpha^*$, which stands for $\alpha$ (or $\log_{10}{\alpha}$) when we use a uniform linear (or log) prior for $\alpha$. 

\begin{figure}
    \centering
    \includegraphics[width=\textwidth]{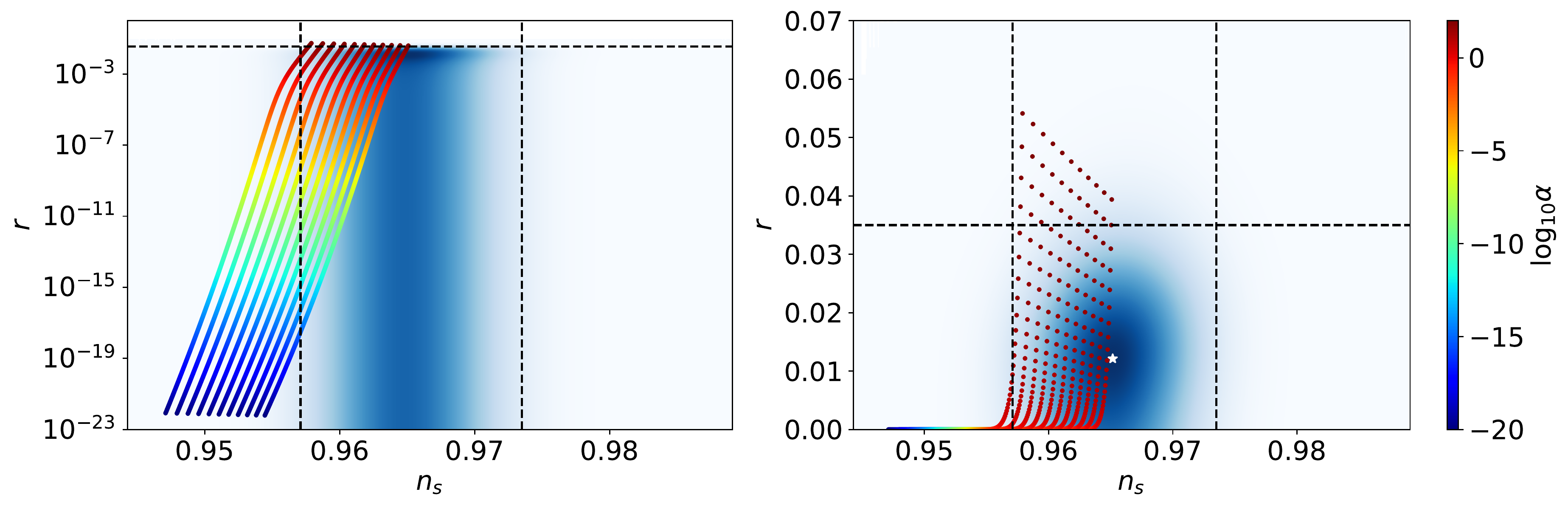}
    \caption{\textit{Planck}+BICEP/Keck 2018 likelihood in the $(n_s,\,r)$ plane shown in blue,
    with
    a logarithmic (linear) scale in the left (right) panel. The black dashed lines indicate the $95\%$ C.L. limits on $n_s$ \eqref{ns planck+BK18} and $r$ \eqref{r bound planck+BK18}, and we highlight in the right panel the maximum likelihood point with a white star. We include the predictions for $n_s$ and $r$ for T-models \eqref{single field potential} with $p=2$, for different values of $\alpha$ (see the colored bar legend), and duration of reheating in the range $0\leq \Delta \tilde N_\text{rh}\leq \Delta \tilde N_\text{rh,max}(\alpha)$. We calculate these using the improved extended predictions, see Eqs.\eqref{ns new p dependence}-\eqref{r new p dependence}.}
    \label{fig:Planck likelihood}
\end{figure}
In Figure \ref{fig:Planck likelihood} we show the $(n_s,\,r)$ likelihood in blue, together with the improved extended predictions, Eqs.\eqref{ns new p dependence}-\eqref{r new p dependence}. Since we are going to use a uniform log (linear) prior on $\alpha$, we employ in the left (right) panel a logarithmic (linear) scale for $r$.  For decreasing $\alpha$, the $\alpha$-dependence in $n_s$ moves the predictions towards the left, into regions with lower likelihood. We note that none of the predictions, i.e. no combination of the inflationary parameters $(\alpha^*, \, \Delta \tilde N_\text{rh})$, exactly coincides with the maximum likelihood point, $(n_{s,\text{max}}, \, r_\text{max})$, marked with a white star in the right plot. The model whose predictions are the closest to the maximum likelihood point is one with $\Delta \tilde N_\text{rh}=0$ (since $(n_{s,\text{max}}, \, r_\text{max})$ lies slightly to the right of the predictions with $\Delta \tilde N_\text{rh}=0$) and $\alpha_\text{max}=3.55$, obtained by solving $r(\alpha_\text{max}, 0)=r_\text{max}$ using Eq.~\eqref{r new p dependence} and $p=2$.

\begin{figure}
\centering
\captionsetup[subfigure]{justification=centering}
   \begin{subfigure}[b]{0.48\textwidth}
    \includegraphics[width=\textwidth]{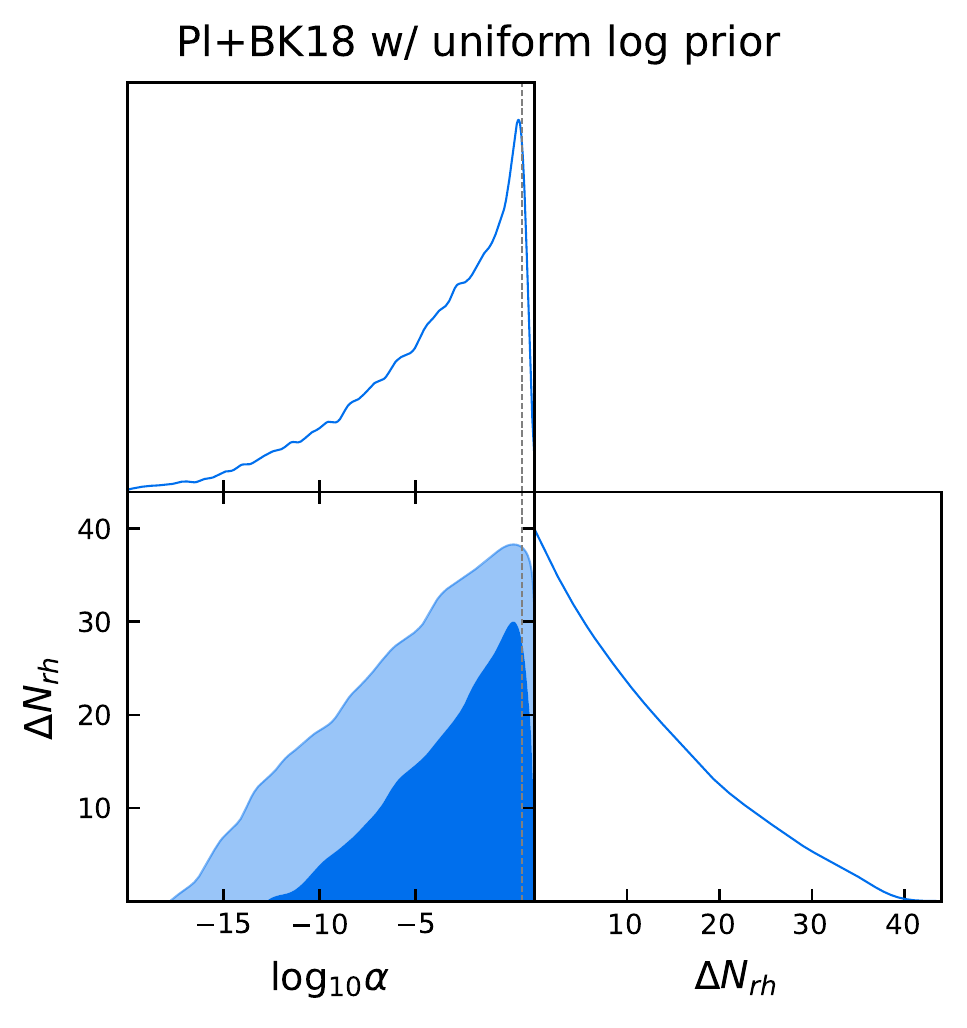}
  \end{subfigure}
   \begin{subfigure}[b]{0.48\textwidth}
    \includegraphics[width=\textwidth]{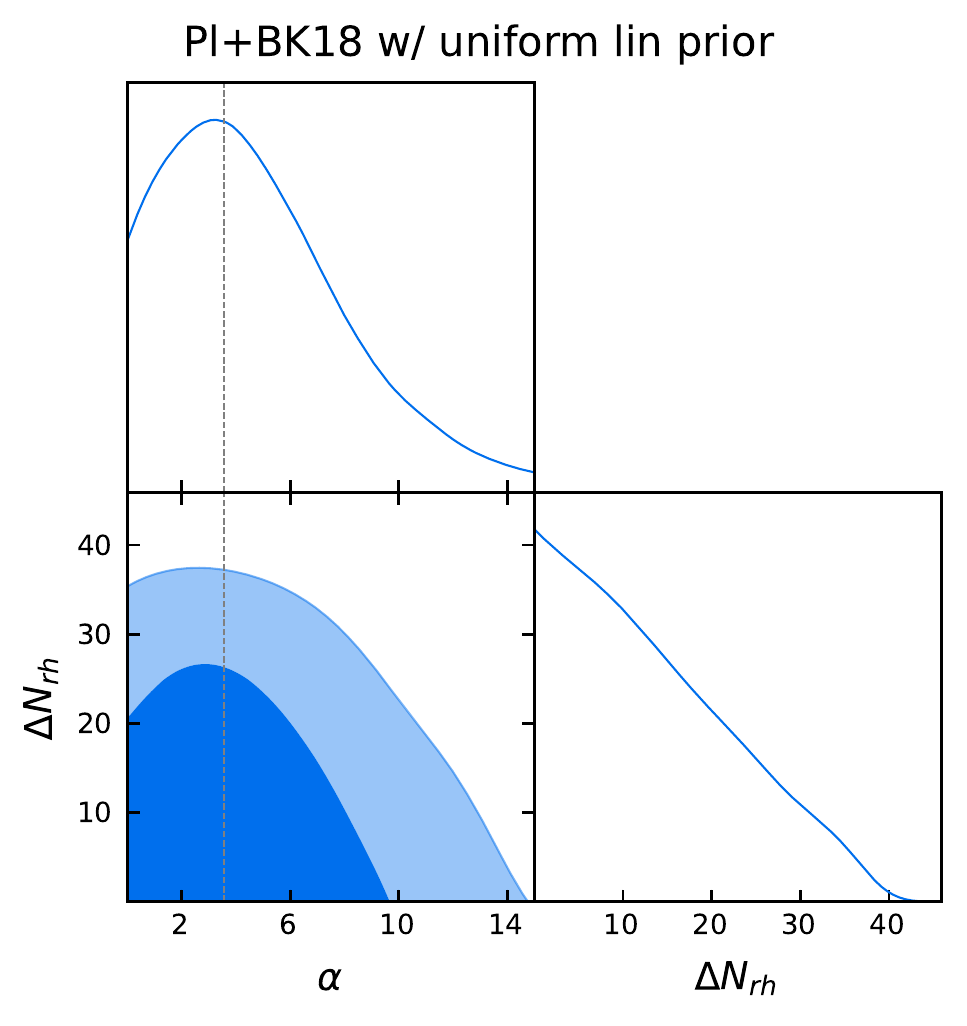}
  \end{subfigure}
\caption{Posterior for the $\alpha^*$ and $\Delta \tilde N_\text{rh}$ inflationary parameters and associated marginalised posteriors obtained using \textit{Planck}+BICEP/Keck 2018 data and the improved extended predictions, see Eqs.\eqref{ns new p dependence}-\eqref{r new p dependence}. The results obtained using a uniform log and uniform linear prior on $\alpha$ are shown in the left and right panels respectively. In each panel, the vertical, dashed line marks $\alpha_\text{max}$, see text for its definition.}
\label{fig:Pl+BK18 posteriors extended}
\end{figure}
We show in Figure \ref{fig:Pl+BK18 posteriors extended} the full posterior for $(\alpha^*, \, \Delta \tilde N_\text{rh})$ and marginalised posterior distributions, on the left (right) the results obtained employing a uniform log (linear) prior on $\alpha$. We find respectively 
\begin{align}
    \label{PlBK18 log prior bounds extended}
    &\text{uniform log prior:}\quad \log_{10}{\alpha} = -4.2^{+5.4}_{-8.6}\; (95\%\text{C.L.})\;, \quad \Delta \tilde N_\text{rh} < 29.6 \; (95\%\text{C.L.})\;,\\
    \label{PlBK18 lin prior bounds extended}
    &\text{uniform linear prior:}\quad \alpha < 11.2\; (95\%\text{C.L.})\;, \quad \; \Delta \tilde N_\text{rh} < 32.5\; (95\%\text{C.L.})\;.
\end{align}
The analysis performed employing the uniform log prior on $\alpha$ shows that $\alpha$ and $\Delta \tilde N_\text{rh}$ are correlated, i.e. small $\alpha$ values favour short reheating stages. When $\alpha$ becomes of order unity the allowed duration of reheating is maximised and then it decreases for larger $\alpha$. 
The marginalised posterior for $\alpha$ peaks close to the value associated with the maximum likelihood point, $\alpha_\text{max}$, highlighted with a dashed line in Figure \ref{fig:Pl+BK18 posteriors extended}. The behavior of the posterior can be readily understood using the left panel of Figure~\ref{fig:Planck likelihood}; the lower bound on $n_s$ combined with the $\alpha$-dependence of $n_s$, see Eq.~\eqref{ns new p dependence} (or equivalently the slope of the predictions in the $(n_s,\,r)$ plane, see left panel of Figure \ref{fig:Planck likelihood}), leads to a lower bound on the value of $\alpha$. To our knowledge, the result in \eqref{PlBK18 log prior bounds extended} is the first lower bound reported on the magnitude of $\alpha$, i.e., on $\log_{10}{\alpha}$, for $\alpha$-attractor T-models with $p=2$.

When using a uniform linear prior on $\alpha$, we find an upper bound on $\alpha$, but no lower bound.
Interestingly, the marginalised posterior peaks close to $\alpha_\text{max}$ and decreases for smaller $\alpha$. This can be understood by looking at the right panel of Figure \ref{fig:Planck likelihood}, where we see that the upper bound on $\alpha$ is driven by the upper bound on $r$, and the predictions for small $\alpha$ move to the left of the $(n_s,\,r)$ plane, due to the $\alpha$-dependence of $n_s$, towards regions with lower likelihood. 

In both cases we see the effect of the $\alpha$-dependence of the scalar spectral tilt. This is of particular importance when using a uniform log prior on $\alpha$, as it leads to a lower bound on $\log_{10}{\alpha}$, see \eqref{PlBK18 log prior bounds extended}. When we employ the uniform linear prior, the effect can be seen in the fact that the posterior decreases for $\alpha$ small, but in this case there is no lower bound on $\alpha$ at 95\% C.L., see \eqref{PlBK18 lin prior bounds extended}. 

Using either prior for $\alpha$, we obtain an upper bound on the duration of reheating, see \eqref{PlBK18 log prior bounds extended}-\eqref{PlBK18 lin prior bounds extended}. While the bounds on $\alpha$ are sensitive to the choice of the prior, we note that the upper $95\%$ C.L. limit on the duration of reheating does not depend strongly on the choice of prior for $\alpha$.

\begin{figure}
    \centering
    \includegraphics[width=.5\textwidth]{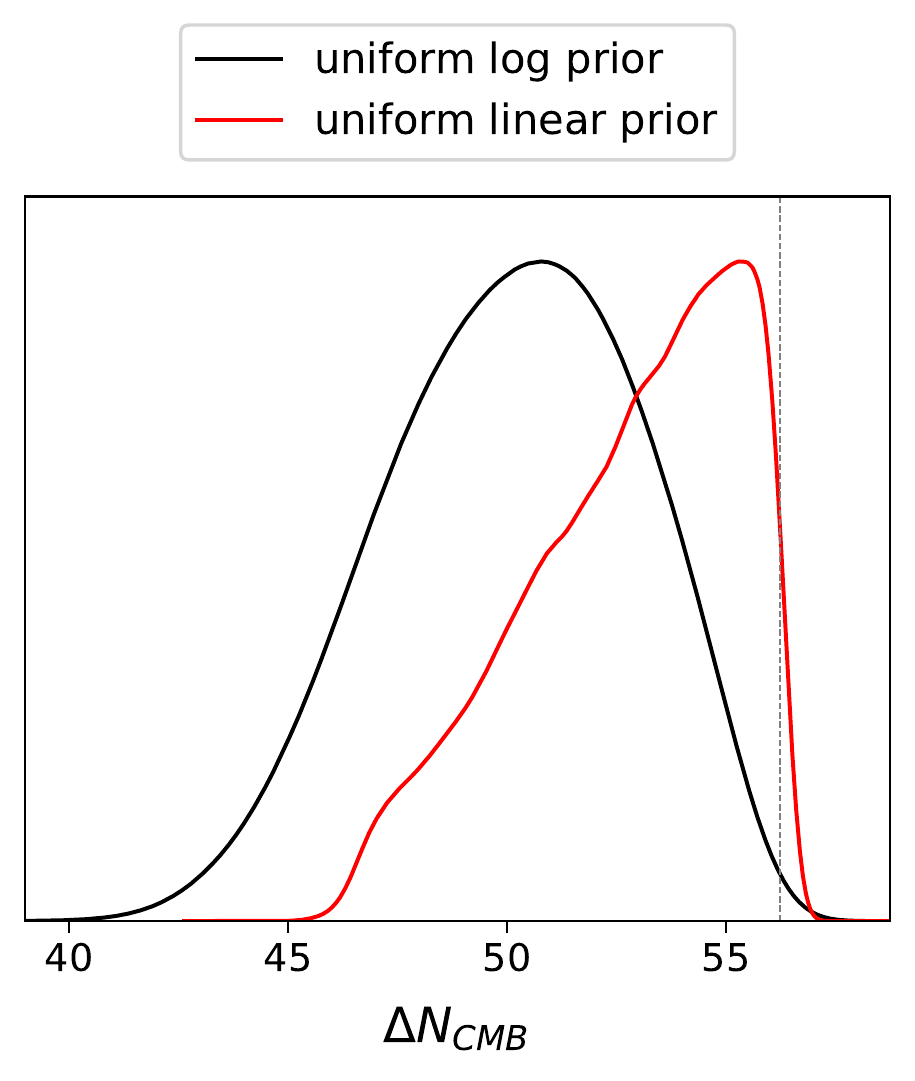}
    \caption{Posterior distribution for $\Delta N_\text{CMB}$, derived from the $\alpha^*$ and $\Delta \tilde N_\text{rh}$ MCMC chains obtained when using the improved extended predictions, Eqs.\eqref{ns new p dependence}-\eqref{r new p dependence}. The results obtained using a uniform log  and uniform linear prior on $\alpha$ are represented with a black and red line respectively. We highlight with a vertical, dashed line $\Delta N_\text{CMB}(\alpha_\text{max},\,0)$, i.e. the value corresponding to the model that is closest to the maximum likelihood point, see Eq.\eqref{N CMB with rh}.}
    \label{fig:posterior Delta NCMB}
\end{figure}
In Figure \ref{fig:posterior Delta NCMB} we show the posterior distribution for $\Delta N_\text{CMB}$, derived using Eq.\eqref{N CMB with rh} and the MCMC chains for $\alpha^*$ and $\Delta \tilde N_\text{rh}$. We find
\begin{align}
    \label{PlBK18 NCMB log prior bounds}
    &\text{uniform log prior:}\quad \Delta N_\text{CMB} = 50.1^{+5.3}_{-5.6} \quad  (95\%\text{C.L.})\;,\\
    \label{PlBK18 NCMB lin prior bounds}
    &\text{uniform linear prior:}\quad \Delta N_\text{CMB}= 52.8^{+3.7}_{-4.9} \quad (95\%\text{C.L.})\;.
\end{align}
The posteriors peak at different values, with the log-posterior preferring smaller $\Delta N_\text{CMB}$. 
The shape of the posterior in the linear-prior case displays a cut-off for large $\Delta N_\text{CMB}$ values. This is due to the prior range used for $\alpha$; $\Delta N_\text{CMB}$ increases for larger $\alpha$ values, see Figure \ref{fig:N CMB}, and for fixed $\alpha$ is maximised in the case of instant reheating, see Eq.\eqref{N CMB with rh}, therefore $\Delta N_\text{CMB}>56.8$ would require models with $\alpha>15$, excluded by the prior. 

We have also considered the case of the improved universal predictions, see Eqs.\eqref{ns new universal}-\eqref{r new universal}. 
Importantly, we find that the posterior distribution qualitatively has the same shape as in Figure \ref{fig:Pl+BK18 posteriors extended}. This clarifies \textit{which} $\alpha$-dependence in the $n_s$ improved extended predictions \eqref{ns new p dependence} is responsible for the lower bound on $\log_{10}\alpha$. Indeed, understanding this point might be obscured by the fact that in Eq.\eqref{ns new p dependence} $\Delta N_\text{CMB}$ is a function of $\alpha$ \textit{and} there are some explicitly $\alpha$-dependent terms (that can be neglected in the large-$\Delta N_\text{CMB}$ limit). On the other hand, the novel $\alpha$-dependence in the universal prediction \eqref{ns new universal} is only due to $\Delta N_\text{CMB}(\alpha, \, \Delta\tilde N_\text{rh})$. Since we observe the same qualitative behavior in both cases, we can establish that it is the $\alpha$-dependence of $\Delta N_\text{CMB}$ that gives rise to a lower bound on the magnitude of $\alpha$. 


\subsection{Standard universal predictions}
\label{sec: naive universal predictions bayesian analysis}
When employing the standard universal predictions \eqref{ns universal}-\eqref{r universal} the inflation parameters are $\alpha^*$ and $\Delta N_\text{CMB}$. In this case there is no dependence of $\Delta N_\text{CMB}$ on either $\alpha$ or on the duration of reheating; $\Delta N_\text{CMB}$ is treated as an independent parameter. We use the same prior definition as in section \ref{sec:improved extended predictions bayesian analysis} for $\alpha^*$, and employ a flat, linear prior on $\Delta N_\text{CMB}$ over the range $\Delta N_\text{CMB}\in [35,\,85]$. The range selected is as permissive as possible, such that theoretical priors on $\Delta N_\text{CMB}$ derived in the literature for other models are not incorrectly applied here, i.e. we let the data decide.

\begin{figure}
\centering
\captionsetup[subfigure]{justification=centering}
   \begin{subfigure}[b]{0.48\textwidth}
    \includegraphics[width=\textwidth]{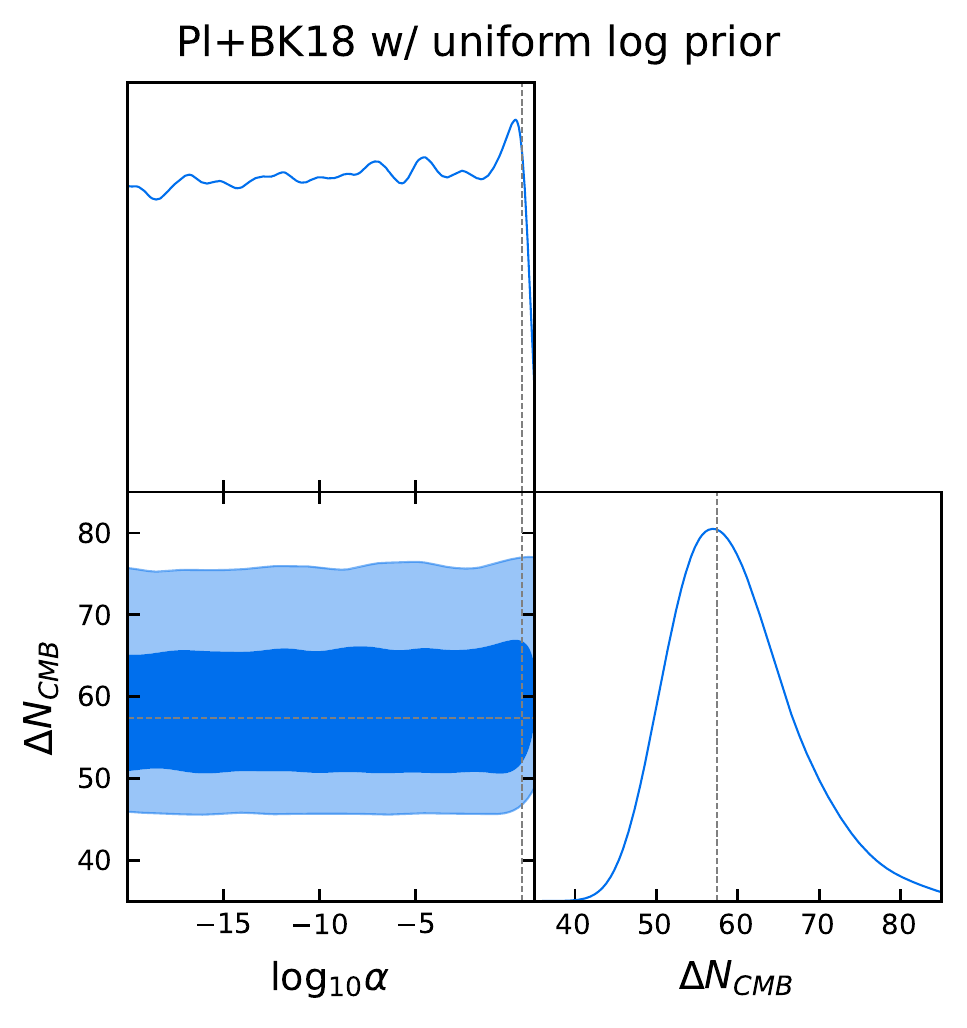}
  \end{subfigure}
   \begin{subfigure}[b]{0.48\textwidth}
    \includegraphics[width=\textwidth]{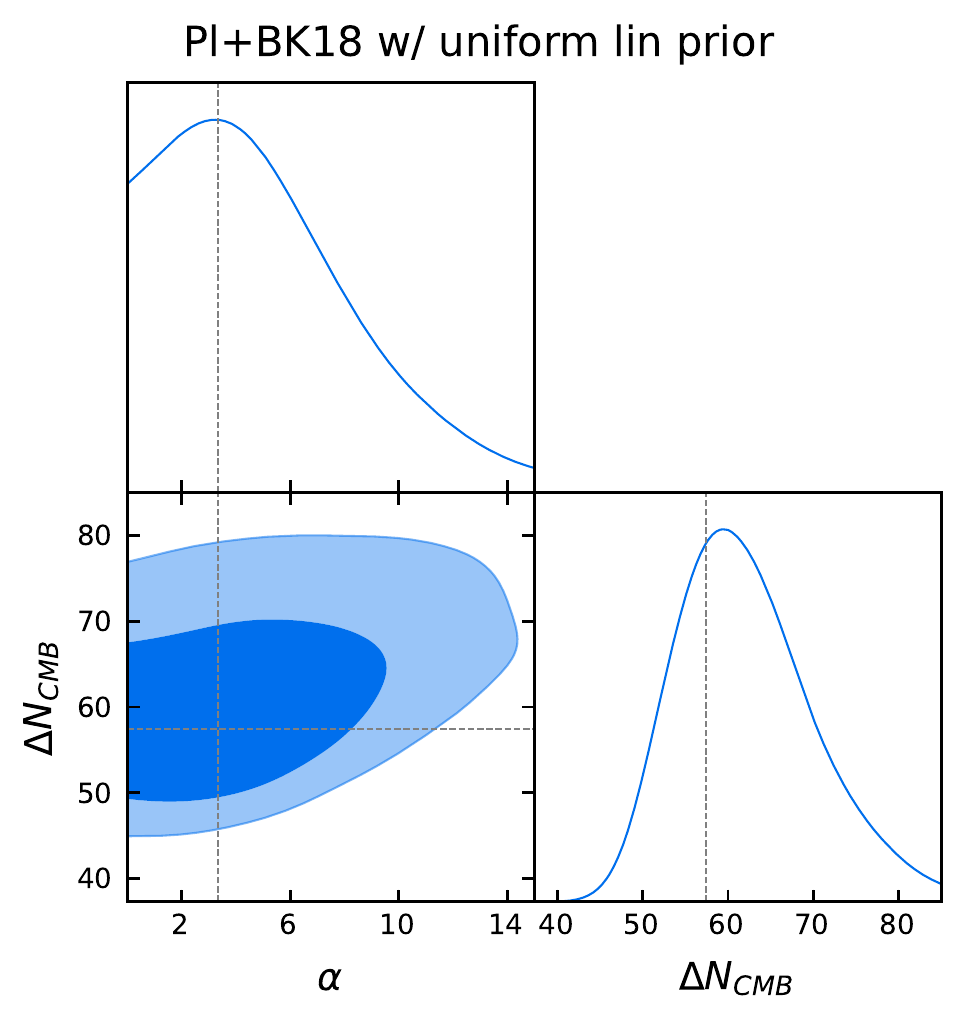}
  \end{subfigure}
\caption{Posterior for the $\alpha^*$ and $\Delta N_\text{CMB}$ inflationary parameters and associated marginalised posteriors obtained using \textit{Planck}+BICEP/Keck 2018 data and the standard universal predictions \eqref{ns universal}-\eqref{r universal}. The results obtained using a uniform log and uniform linear prior on $\alpha$ are shown in the left and right panels respectively. In each panel, the dashed lines mark the $\alpha^*$ and $\Delta N_\text{CMB}$ values that correspond to the maximum likelihood point.}
\label{fig:Pl+BK18 posteriors naive}
\end{figure}
We show in Figure \ref{fig:Pl+BK18 posteriors naive} the full posterior for $(\alpha^*, \, \Delta N_\text{CMB})$ and marginalised posterior distributions, on the left (right) the results obtained employing a uniform log (linear) prior on $\alpha$. We find respectively 
\begin{align}
    \label{PlBK18 log prior bounds naive}
    &\text{uniform log prior:}\quad \log_{10}{\alpha} = \text{not defined}\;, \quad \Delta N_\text{CMB} = 60^{+20}_{-10} \; (95\%\text{C.L.})\;,\\
    \label{PlBK18 lin prior bounds naive}
    &\text{uniform linear prior:}\quad \alpha < 11.5\; (95\%\text{C.L.})\;, \quad \; \Delta N_\text{CMB} = 62^{+20}_{-10}\; (95\%\text{C.L.})\;,
\end{align}
where, due to the flatness of the $\log_{10}{\alpha}$ posterior, the $95\%$ C.L. constraint is not well defined and would depend on the chosen lower end of the logarithmic prior. By comparing these results with those shown in Figure \ref{fig:Pl+BK18 posteriors extended}, one immediately realises that analysing the data using $\Delta N_\text{CMB}$ as an independent inflationary parameter hides the impact of the duration of reheating on observable quantities and the implicit dependence of $\Delta N_\text{CMB}$ on $\alpha$. The first consequence of this is that one cannot derive bounds on the duration of reheating within this analysis, while this is possible when employing the improved extended predictions, see \eqref{PlBK18 log prior bounds extended} and \eqref{PlBK18 lin prior bounds extended}. Moreover, when assuming a uniform log prior on $\alpha$, the standard universal predictions do not yield a lower bound on $\log_{10}{\alpha}$, due to the loss of information caused by not including the implicit $\alpha$-dependence in $\Delta N_\text{CMB}$. This can be seen by comparing the shapes of the marginalised posteriors for $\log_{10}{\alpha}$ in Figures \ref{fig:Pl+BK18 posteriors extended} and \ref{fig:Pl+BK18 posteriors naive}. In the linear prior case, the $95\%$ C.L. upper bound on $\alpha$, see \eqref{PlBK18 lin prior bounds naive}, is very similar to the one derived using the improved predictions, see \eqref{PlBK18 lin prior bounds extended}. 

The marginalised posteriors for $\Delta N_\text{CMB}$ peak close to the value corresponding to the maximum likelihood point, and decrease for very large or very small values.   For these models an extended reheating stage with $w<1/3$ always decreases the value of $\Delta N_\text{CMB}$, see Eq.~\eqref{Nstar}, therefore large\footnote{See e.g. \cite{German:2022sjd}, where the authors derive a model independent expression for $\Delta N_\text{CMB}$, $\Delta N_\text{CMB} = 56.9+\frac{1}{4}\ln{(r)} - \frac{1-3w}{3(1+w)}\ln{({T_\text{end}}/{T_\text{th}})}$, where $T_\text{end}$ ($T_\text{rh}$) is the temperature at the end of inflation (reheating). For an extended reheating stage, $T_\text{th}< T_\text{end}$, reheating contributes positively to $\Delta N_\text{CMB}$ if $w>1/3$. Moreover, by noting that the tensor-to-scalar ratio scales approximately linearly with $\alpha$ for $\alpha$-attractors (see the right panel of Figure \ref{fig:ns and r universal predictions}), we recognise in the logarithmic dependence of $\Delta N_\text{CMB}$ on $r$ found in \cite{German:2022sjd} the logarithmic dependence on $\alpha$ that we find, see Figure \ref{fig:N CMB}.} $\Delta N_\text{CMB}$ values can be produced only within exotic reheating scenarios with $w>1/3$. Large values of $\Delta N_\text{CMB}$ are nevertheless disfavoured, as they are associated with values of $n_s$ and $r$ with low 
likelihood.
The $95\%$ C.L. limits on $\Delta N_\text{CMB}$ do not depend strongly on the choice of prior for $\alpha$.

\section[\texorpdfstring{T-models with $\bm{p=4}$}{T-models with p=4}]{T-models with $\bm{p=4}$}
\label{sec:models with p=4}
As discussed in section \ref{sec: inflation and reheating}, we consider contributions to the potential $V(Z,\,\bar Z)$ proportional to powers of $Z\bar Z$, see Eq.~\eqref{V of Z}, and assume that 
only the leading non-zero term
in the series dominates over the others. For this reason, after analysing the case with $p=2$, we now turn to 
T-models \eqref{single field potential}
with $p=4$. In this section we will show that the $\alpha$-dependence of the scalar spectral tilt for small $\alpha$, as well as its consequences for the $\alpha$ posterior, is robust against a different choice for $p$. 

\begin{figure}
    \centering
    \includegraphics[width=.6\textwidth]{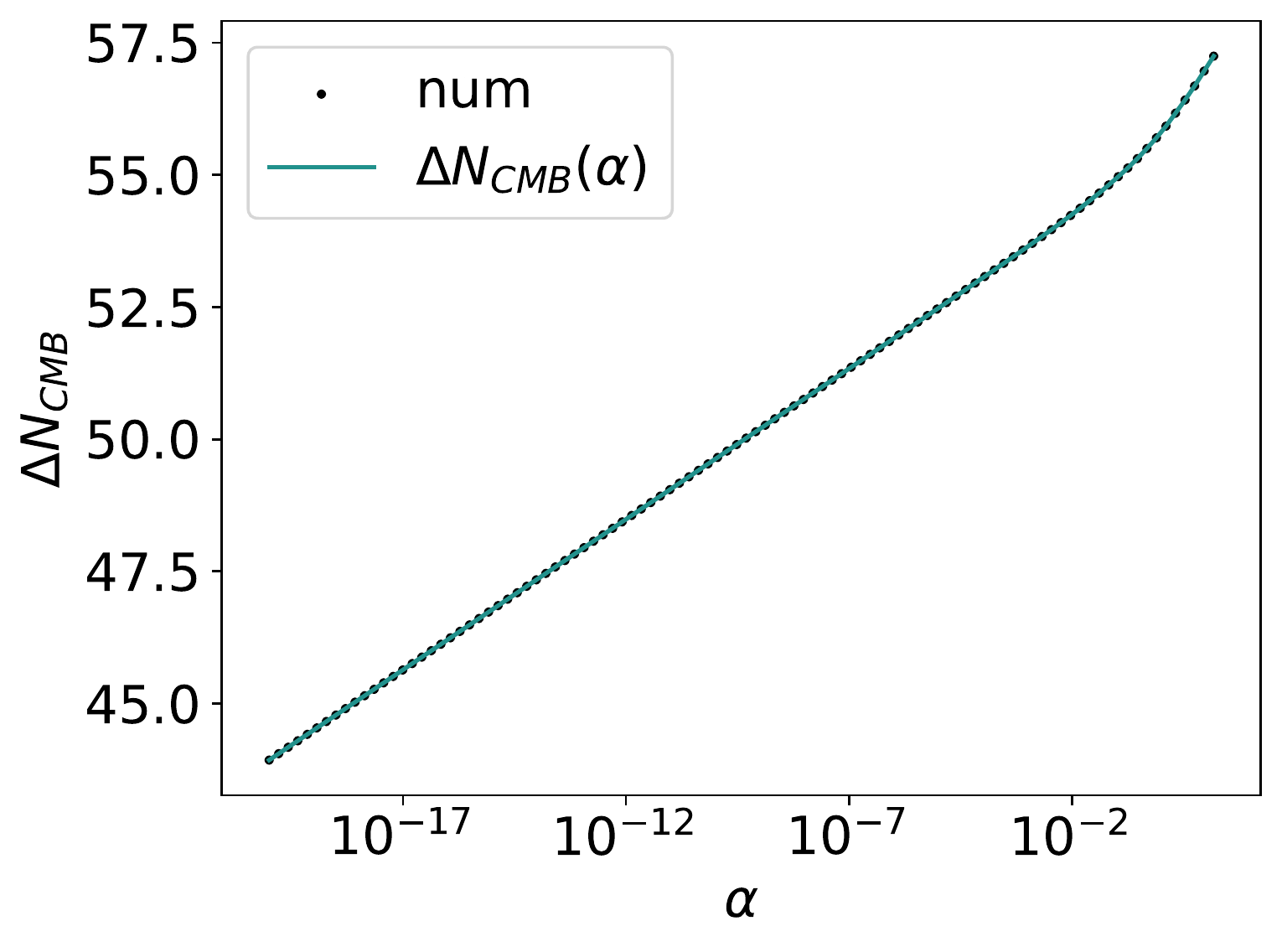}
    \caption{Numerical $\Delta N_\text{CMB}$ values (black dots) for $\alpha$-attractor T-models with potential \eqref{single field potential}, $p=4$ and $10^{-20}\leq \alpha\leq 15$. The green line represents the interpolating function $\Delta N_\text{CMB}(\alpha)$.}
    \label{fig:NCMB p=4}
\end{figure}
We first note an important difference from the case $p=2$, as $\Delta N_\text{CMB}$ is independent of the duration of reheating, see Eq.~\eqref{Nstar}, since Eq.~\eqref{equation of state parameter} gives $w=1/3$ for $p=4$. 
As a result it is not possible to derive constraints on the duration of reheating, as the observables, $n_s$ and $r$, do not depend on $\Delta \tilde N_\text{rh}$.

In Figure~\ref{fig:NCMB p=4} we show the numerical values of $\Delta N_\text{CMB}$ against $\alpha$ for T-models with potential \eqref{single field potential}, $p=4$ and $10^{-20}\leq \alpha\leq 15$. The numerical results are obtained by iteratively solving \eqref{Nstar} for $\Delta N_\text{CMB}$, with values of $V_0$ compatible with the observed scalar power spectrum \eqref{P zeta power law} at the scale $k_\text{CMB}$.
These results show that $\Delta N_\text{CMB}$ scales logarithmically with $\alpha$, as observed for models with $p=2$, see Figure \ref{fig:N CMB}. The analytical derivation of $\Delta N_\text{CMB}(\alpha)$, as was done in section \ref{sec: analytic Delta N CMB} for $p=2$, is more involved for $p=4$. For this reason, we choose to describe the $\alpha$-dependence in $\Delta N_\text{CMB}$ by using an interpolating function, which we represent with a green line in Figure \ref{fig:NCMB p=4}. 

\begin{figure}
    \centering
    \includegraphics[width=\textwidth]{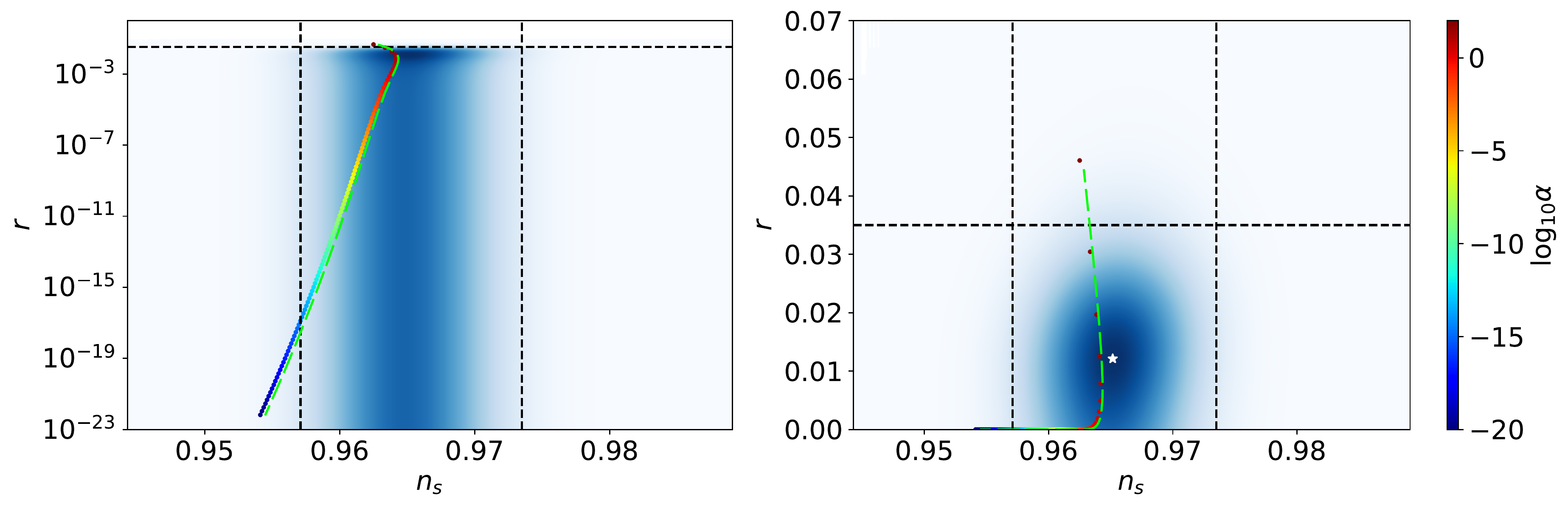}
    \caption{\textit{Planck}+BICEP/Keck 2018 likelihood in the $(n_s,\,r)$ plane. We represent the same figure, employing a logarithmic (linear) scale on the left (right). We mark with dashed, black lines the $95\%$ C.L. limits on $n_s$ and $r$, and highlight in the right panel the maximum likelihood point with a white star. 
    The colored points represent the numerical predictions for T-models \eqref{single field potential} with $p=4$ and different values of $\alpha$, see the colored bar legend. They are compared with the improved extended predictions (green, dashed line), which we obtain from Eqs.\eqref{ns p dependence}-\eqref{r p dependence} by using the interpolating function $\Delta N_\text{CMB}(\alpha)$, see Figure \ref{fig:NCMB p=4}, and setting $p=4$.}
    \label{fig:ns and r p=4}
\end{figure}
In Figure \ref{fig:ns and r p=4} we show the 
predictions for $n_s$ and $r$, obtained by numerically evaluating $n_s$ at second-order in slow-roll, see Eq.~\eqref{ns second order HSRP}, and $r$ at first-order, see Eq.~\eqref{r first order HSRP}, together with the \textit{Planck}+BICEP/Keck 2018 likelihood in the $(n_s,\,r)$ plane. The left and right panels differ only by the log or linear scale chosen for the vertical axis. For large $\alpha$, the numerical predictions display a turn in the $(n_s, \, r)$ plane, showing a different behavior with respect to the $p=2$ case, see Figure \ref{fig:ns r slow roll}. We compare these results with the improved extended predictions, obtained from Eqs.\eqref{ns p dependence}-\eqref{r p dependence} by setting $p=4$ and using the interpolating function $\Delta N_\text{CMB}(\alpha)$. These are represented with a green, dashed line in Figure \ref{fig:ns and r p=4}, showing a very good agreement with the numerical results. In analogy with what done in section \ref{sec:improved predictions} for $p=2$, we checked that $|n_{s,\text{num}}-n_{s,\text{extended}}|<\sigma(n_s)$ and $|r_\text{num} -r_\text{extended}|<\sigma(r)$ for the range of $\alpha$ considered. The improved extended predictions can be therefore safely used when comparing these models with \textit{Planck} and BICEP/Keck 2018 data.

Except for small deviations for large $\alpha$, $\Delta N_\text{CMB}(\alpha)$ and the predictions for $n_s$ and $r$ are almost identical to those for models with $p=2$ and instant reheating. The fact that these results coincide for small $\alpha$ is consistent with the expected $\alpha$-attractor behaviour. On the other hand, this characteristic behaviour of $\alpha$-attractors gets washed out for large $\alpha$, where these models approach the simple chaotic inflation behaviour \cite{Kallosh:2015zsa}. 

\begin{figure}
\begin{subfigure}{.5\textwidth}
  \centering
  \includegraphics[width=\textwidth]{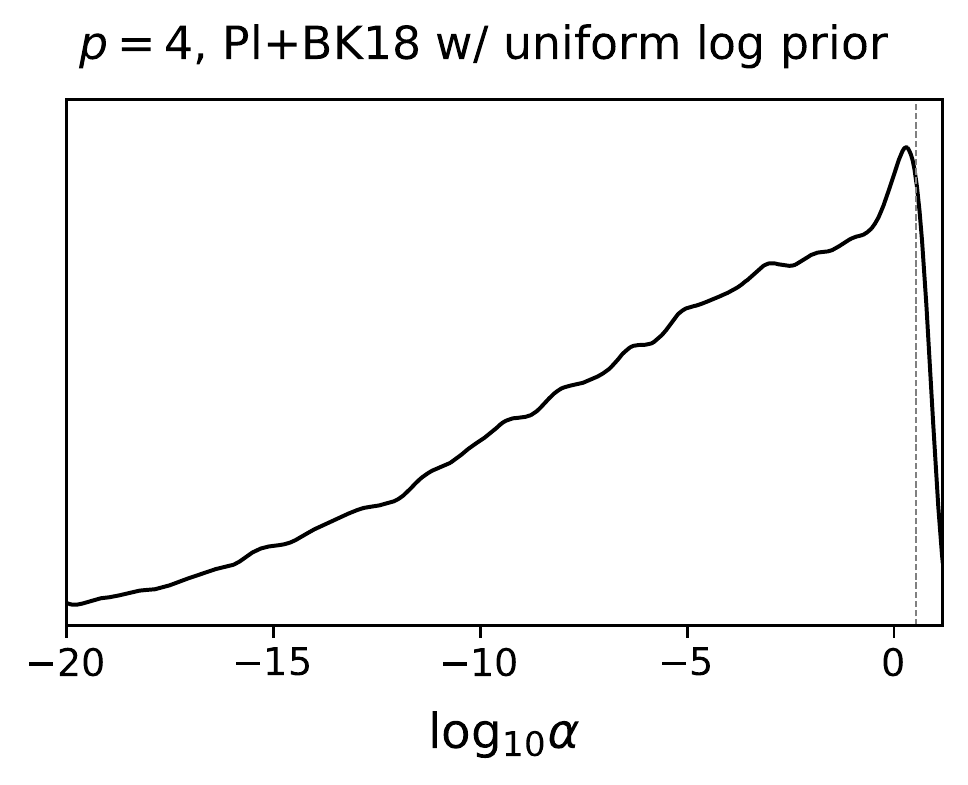}
\end{subfigure}
\begin{subfigure}{.5\textwidth}
  \centering
  \includegraphics[width=\textwidth]{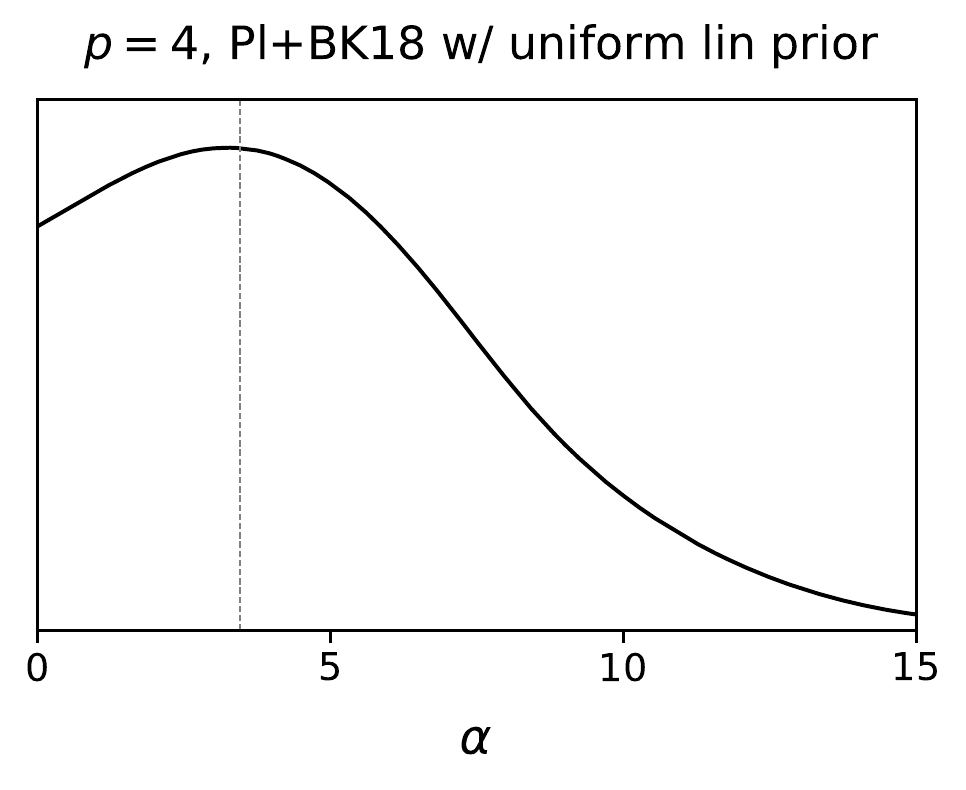}
\end{subfigure}
\caption{Posterior for $\log_{10}{\alpha}$ (left panel) and $\alpha$ (right panel) obtained using \textit{Planck}+BICEP/Keck 2018 data for T-models \eqref{single field potential} with $p=4$. The results on the left (right) are obtained using a uniform log (linear) prior on $\alpha$. In each panel, the vertical, dashed line marks $\alpha_\text{max}$, see text for its definition.}
\label{fig:posterior p=4}
\end{figure}
We show in Figure \ref{fig:posterior p=4} the posterior for $\log_{10}{\alpha}$ (left panel) and $\alpha$ (right panel), obtained employing a uniform log and linear prior respectively. We define the $\alpha$ priors as in section \ref{sec:improved extended predictions bayesian analysis}. We find 
\begin{align}
    \label{PlBK18 log prior bounds naive p=4}
    &\text{uniform log prior:}\quad \log_{10}{\alpha} = -5.7^{+6.8}_{-9.5}\; (95\%\text{C.L.})\;,\\
    \label{PlBK18 lin prior bounds naive p=4}
    &\text{uniform linear prior:}\quad \alpha  < 10.8\; (95\%\text{C.L.})\;.
\end{align}
The posteriors peak close to $\alpha_\text{max}$, the $\alpha$ value whose predictions are the closest to the maximum likelihood point. The shape of the posteriors is similar to that recovered in the $p=2$ case, see Figure \ref{fig:Pl+BK18 posteriors extended}. In particular, the $\alpha$-dependence in the scalar spectral tilt yields a lower bound on $\log_{10}{\alpha}$ also in this case, implying that the existence of a lower bound on $\log_{10}\alpha$ is robust against these two different choices of $p$. 

For a generic potential with a non-zero mass term, see Eq.\eqref{V of Z}, we expect that any non-zero quadratic term eventually dominates at small inflaton-field values, yielding $w=0$. For this reason, we analyse T-models with $p=4$ including an extended reheating stage with $w=0$. We define $\Delta N_\text{CMB}(\alpha, \, \Delta \tilde N_\text{rh})$ according to Eq.\eqref{N CMB with rh}, using the interpolating function represented in Figure \ref{fig:NCMB p=4}. The improved extended predictions are then given by Eqs.\eqref{ns new p dependence}-\eqref{r new p dependence}, where we substitute $p=4$. We define the prior on $\alpha^*$ and on $\Delta \tilde N_\text{rh}$ as in section \ref{sec:improved extended predictions bayesian analysis}, and employ $\Delta \tilde N_\text{rh,max}$ values obtained for models with $p=4$.
\begin{figure}
\begin{subfigure}{.5\textwidth}
  \centering
  \includegraphics[width=\textwidth]{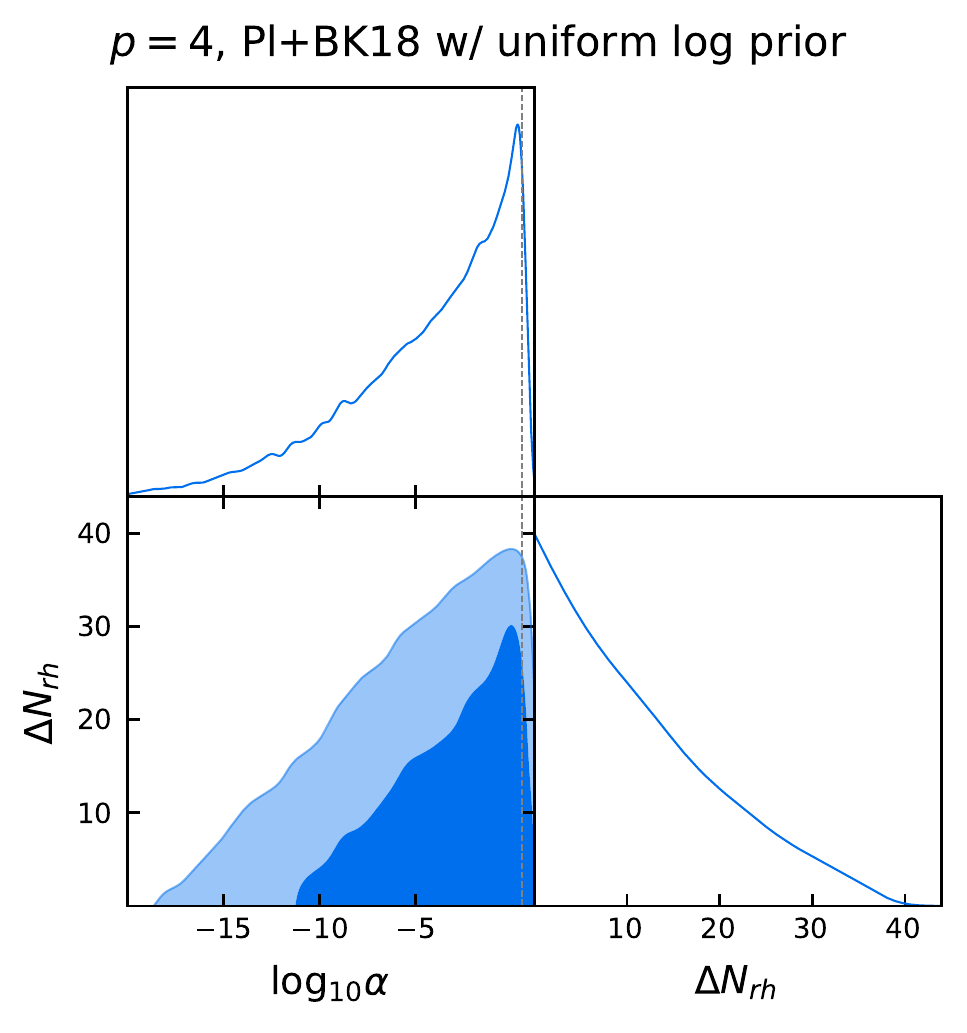}
\end{subfigure}
\begin{subfigure}{.5\textwidth}
  \centering
  \includegraphics[width=\textwidth]{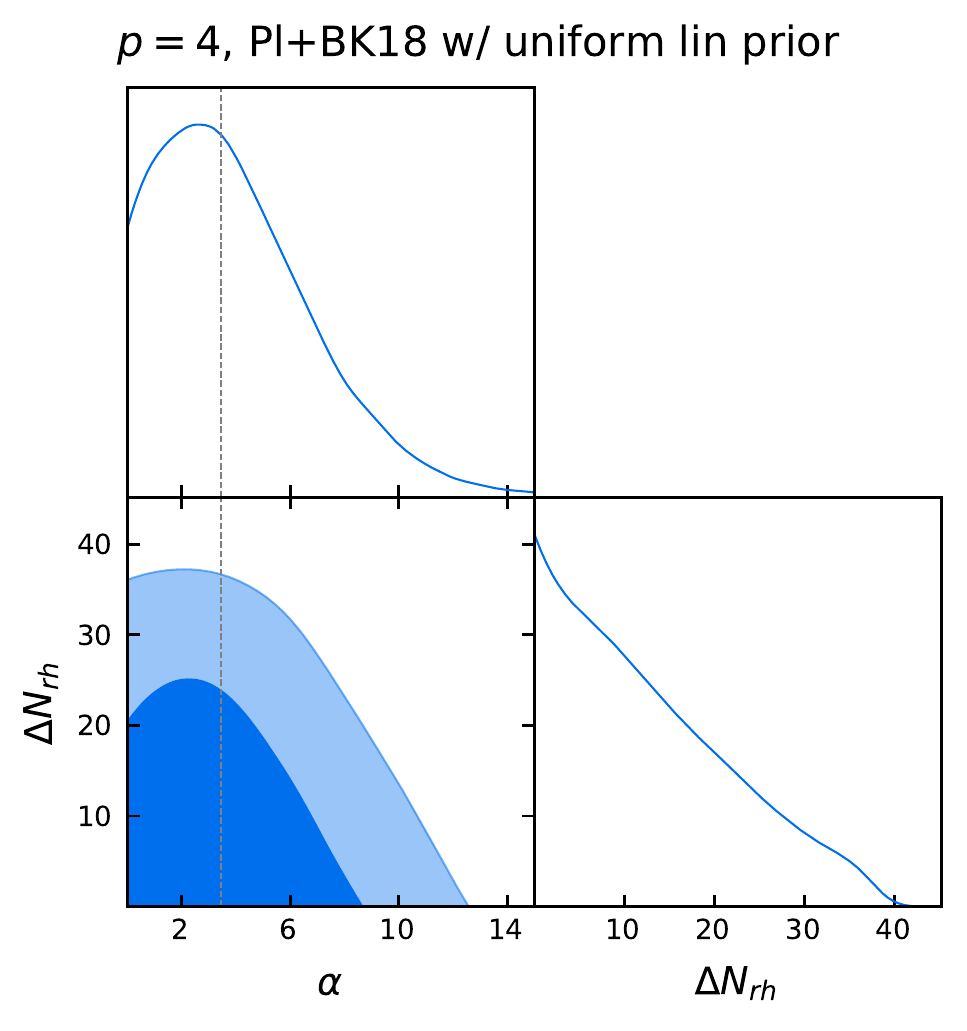}
\end{subfigure}
\caption{Posterior for the $\alpha^*$ and $\Delta \tilde N_\text{rh}$ inflationary parameters and associated marginalised posteriors obtained using \textit{Planck}+BICEP/Keck 2018 data and the improved extended predictions, Eqs.\eqref{ns new p dependence}-\eqref{r new p dependence}, for T-models \eqref{single field potential} with $p=4$. See text for more details. The results on the left (right) are obtained using a uniform log (linear) prior on $\alpha$. In each panel, the vertical, dashed line marks $\alpha_\text{max}$, i.e. the $\alpha$ value of the model whose predictions are the closest to the
maximum likelihood point.}
\label{fig:posterior p=4 with reheating}
\end{figure}
We show in Figure \ref{fig:posterior p=4 with reheating} the full posterior for $(\alpha^*, \,\Delta \tilde N_\text{rh})$ and marginalised posterior distributions, on the left (right) the results obtained employing a uniform log (linear) prior on $\alpha$. We find respectively
\begin{align}
    \label{PlBK18 log prior bounds naive p=4 with rh}
    &\text{uniform log prior:}\quad \log_{10}{\alpha} = -4.3^{+5.4}_{-8.6}\; (95\%\text{C.L.})\;, \quad \Delta \tilde N_\text{rh} < 29.4 \; (95\%\text{C.L.})\;,\\
    \label{PlBK18 lin prior bounds naive p=4 with rh}
    &\text{uniform linear prior:}\quad \alpha  < 9.49\; (95\%\text{C.L.})\;, \quad \Delta \tilde N_\text{rh} < 31.6 \; (95\%\text{C.L.})\;. 
\end{align}
We can compare these results with those obtained for T-models with $p=2$, see \eqref{PlBK18 log prior bounds extended} and \eqref{PlBK18 lin prior bounds extended}. The upper $95\%$ C.L. bound on $\alpha$ is slightly stronger with respect to the $p=2$ case. This is expected considering that the predictions are different for large $\alpha$, see Figures \ref{fig:ns r slow roll} and \ref{fig:ns and r p=4}. On the other hand, the $95\%$ C.L. bounds on $\log_{10}\alpha$ are almost identical. This is due to the fact that the large-scale observables of models with $p=2$ and $p=4$ are the same for small $\alpha$ and $w=0$, see also the discussion in section \ref{sec: perspective on our results}. This result implies that not only the existence of the lower bound on the magnitude of $\alpha$, but also its $95\%$ C.L. value, is robust against these two different choices of the potential profile.

\section{Discussion}
\label{sec: discussion}

\subsection{Perspective on our results}
\label{sec: perspective on our results}
In this work we derive novel CMB constraints on $\alpha$-attractor T-models of inflation. We show that correctly including the $\alpha$-dependence of $\Delta N_\text{CMB}$ yields better analytic predictions for large-scale observables, see Eqs.\eqref{ns new p dependence}-\eqref{r new p dependence}. These \textit{improved} predictions describe accurately the numerical values of $n_s$ and $r$, see Figure \ref{fig:check improved uni predictions}, and also allow us to constrain the duration of reheating from observations.  By working in the Bayesian framework and using both uniform logarithmic and linear priors for $\alpha$, we compare the improved analytic predictions against \textit{Planck} and BICEP/Keck 2018 data. We derive bounds on $\alpha$ (or $\log_{10}\alpha$, when using the uniform log prior on $\alpha$) and the duration of reheating, see \eqref{PlBK18 log prior bounds extended} and \eqref{PlBK18 lin prior bounds extended}. In particular, we show that the $\alpha$-dependence yields a new lower bound on the magnitude of $\alpha$ in these models, see \eqref{PlBK18 log prior bounds extended}.    

The existence of such a bound is conceptually important, and it is not directly evident from the universal predictions alone. Indeed, by inspecting Eqs.\eqref{ns universal} and \eqref{r universal} it might seem that, in absence of detection of primordial tensor modes, $\alpha$-attractor models can be made compatible with CMB constraints for arbitrary small values of $\alpha$. Our results show that this is not the case, and values of $\alpha$ are already bounded from below by current CMB measurements. This implies also that the value of the tensor-to-scalar ratio cannot be arbirtarily small within these models.

Besides being interpreted as a phenomenological parameter connected with the amplitude of primordial tensor modes, see Eq.\eqref{r universal}, the parameter $\alpha$ has a specific meaning in the more fundamental theory from which $\alpha$-attractors are formulated, namely supergravity models. In fact, $\alpha$ is related to the radius of the Poincaré disk, $R=\sqrt{3\alpha}$, on which the complex field $Z$, see Eq.\eqref{kinetic lagrangian 1}, lives \cite{Kallosh:2015zsa}. The smaller $\alpha$, the smaller the disk. Therefore, bounding $\alpha$ with CMB measurements translates into bounds on supergravity theories.

Our work sheds new light on analytical predictions for large-scale observables from $\alpha$-attractor models, taking into account the full dependence on $\alpha$ and on the duration of reheating. We take these to be the fundamental inflationary parameters (as opposed to $\alpha$ and $\Delta N_\text{CMB}$). This can be seen by comparing the standard universal predictions, Eqs.\eqref{ns universal} and \eqref{r universal}, with our improved expressions, Eqs.\eqref{ns new p dependence} and \eqref{r new p dependence}. According to the principles of the Bayesian framework, all prior knowledge should be taken into account. Using the standard universal predictions, Eqs.\eqref{ns universal}-\eqref{r universal}, leads to a loss of information, at odds with the philosophy of the Bayesian approach. As a result the duration of reheating is not directly constrained and there is no lower bound on the magnitude of $\alpha$, see \eqref{PlBK18 log prior bounds naive}-\eqref{PlBK18 lin prior bounds naive}. 

In this work we have explicitly considered T-models with $p=2$ and $p=4$, establishing that the $\alpha$-dependence of the large-scale observables applies to both of these choices of $p$. Our analysis shows that the lower bound on the magnitude of $\alpha$ is robust against these two choices of $p$, see \eqref{PlBK18 log prior bounds extended} and \eqref{PlBK18 log prior bounds naive p=4 with rh}. We note that this holds provided that reheating is matter dominated ($w=0$). For a generic potential \eqref{V of Z} with a non-zero mass term, we expect that any non-zero quadratic term in the potential eventually dominates at small inflaton-field values, yielding $w=0$.
\begin{figure}
    \centering
    \includegraphics[width=.75\textwidth]{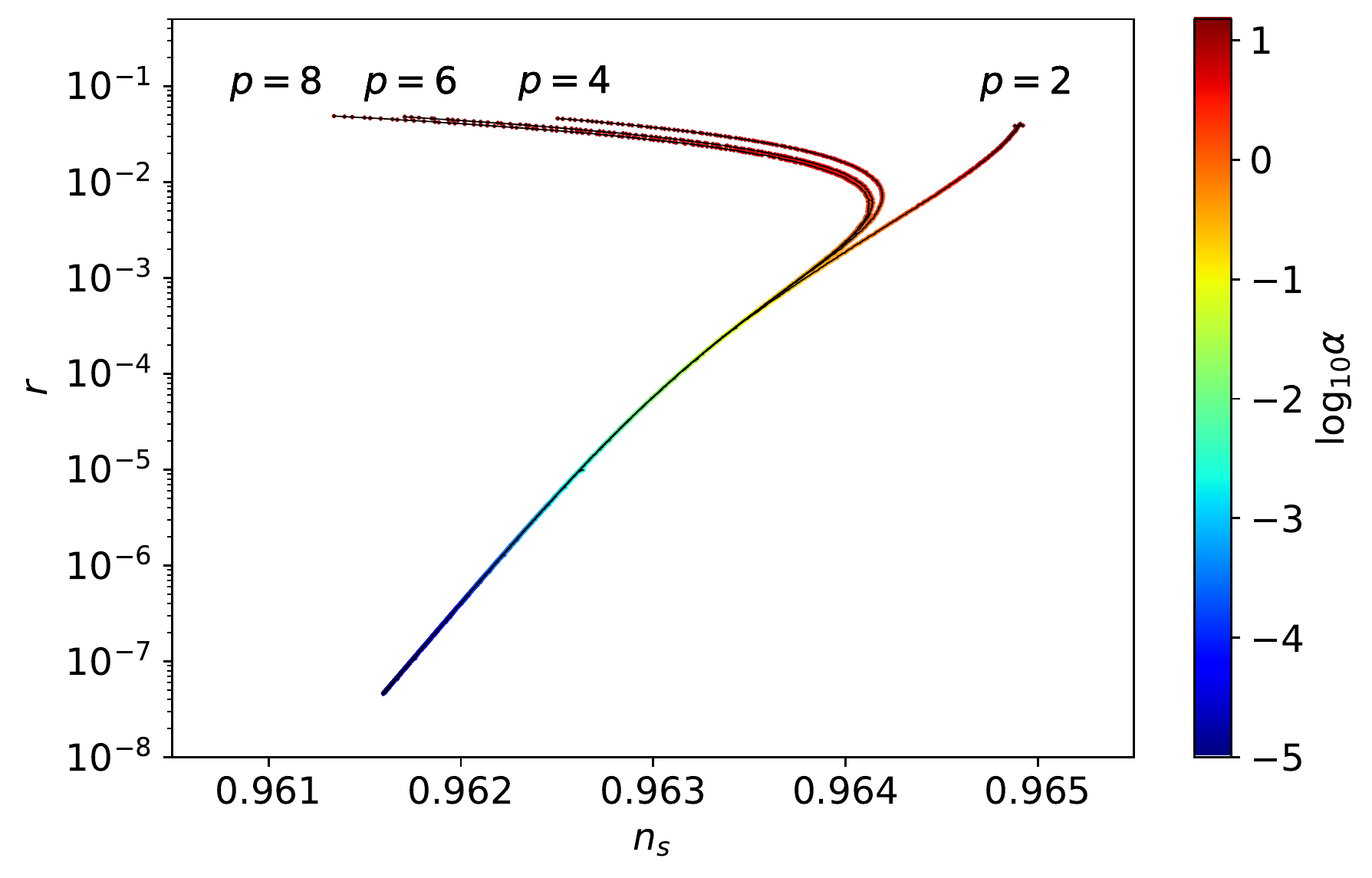}
    \caption{Numerical predictions in the $(n_s,\,r)$ plane for T-models with $p=\{2,\,4,\,6,\,8\}$ and different values of $\alpha$. These results are obtained assuming instant reheating.}
    \label{fig:change p}
\end{figure}
We show in Figure \ref{fig:change p} the numerical predictions for $n_s$ and $r$ for T-models with different values of $p$, calculated assuming instant reheating. The predictions of models with $p>2$ display a turn in the $(n_s,\,r)$ plane, as already pointed out in section \ref{sec:models with p=4} for the $p=4$ case, and converge for increasing $p$. Interestingly, for $\alpha\lesssim 0.1$ they are independent of the $p$ value, consistent with the expected $\alpha$-attractor behavior. We therefore expect that the lower bound on the magnitude of $\alpha$ found for the $p=2$ and $p=4$ cases will be robust against changes in the potential shape, i.e. other choices of $p$. This holds provided that reheating is matter-dominated ($w=0$). In this case, a finite reheating phase moves the predictions towards smaller values of the scalar spectral tilt for fixed $\alpha$ and $p$, and the instant reheating scenario, whose predictions are represented in Figure \ref{fig:change p}, yields the most conservative lower bound on $\log_{10}{\alpha}$. 

\subsection{Comparison with the literature}
\label{sec: compare with literature}

In \cite{LinaresCedeno:2022rbq} the authors perform a Bayesian analysis of a generalisation of $\alpha$-attractor T-models. Their posterior analysis shows a correlation between the parameter $\alpha$ and $\Delta N_\text{CMB}$, which we analytically justify, see Eq.\eqref{N CMB}. Similar to our results, see Figure \ref{fig:ns and r universal predictions}, they also find $n_s$ is correlated with $\alpha$. Interestingly, according to the analysis of \cite{LinaresCedeno:2022rbq}, these $\alpha$-attractors seem to prefer the specific value $r\simeq 0.0025$. We suspect that this result strongly depends on the prior chosen for $\alpha$, which in \cite{LinaresCedeno:2022rbq} is bounded from below due to numerical precision issues and adopts a uniform linear prior. As demonstrated in this work, the constraints on $\alpha$ change depending on the choice of prior, and therefore allowed $r$ values will change accordingly (since $r$ is strongly correlated with $\alpha$, see Figure \ref{fig:ns and r universal predictions}). It would be interesting to explicitly assess the impact of different prior choices in the context of the analysis of \cite{LinaresCedeno:2022rbq}. 

Other works have previously addressed CMB constraints on $\alpha$-attractor models and reheating, see e.g. \cite{Ueno:2016dim, Nozari:2017rta, Drewes:2017fmn, Mishra:2021wkm, Ellis:2021kad, Ling:2021zlj, Drewes:2022nhu, Drewes:2023bbs, Chakraborty:2023ocr}. For example, in \cite{Ellis:2021kad} T-models with $p=2$ are compared with \textit{Planck}, BAO and BICEP+Keck 2018 data. However, there are several differences between \cite{Ellis:2021kad} and our work. In \cite{Ellis:2021kad} the minimum duration of reheating is constrained from the overproduction of supersymmetric dark matter, while we are agnostic about reheating process adopting only a minimal requirement that reheating is completed before the onset of Big Bang nucleosynthesis.
The upper bound on $\alpha$ is obtained for a specific value of the lightest supersymmetric particle mass. The bound, $\alpha<11$ at $95\%$ C.L., is very close to the one we get using the improved analytical predictions and a uniform linear prior, see \eqref{PlBK18 lin prior bounds extended}, even if technically the analysis is different as we derive our constraints within a Bayesian framework. Unlike \cite{Ellis:2021kad} we also consider a uniform logarithmic prior for $\alpha$, exploring the impact of the choice of prior. 

Our \textit{Planck}+BICEP/Keck 2018 constraints on $\log_{10}{\alpha}$ allow for very small $\alpha$ values, see \eqref{PlBK18 log prior bounds extended}. The phenomenology of T-models with $\alpha\simeq \mathcal{O}(10^{-3})$ has recently been considered in \cite{Iacconi:2021ltm}, in the context of multi-field $\alpha$-attractors. The parameter $\alpha$ is inversely proportional to the curvature of the field space, and $\alpha\ll1$ corresponds to strongly-curved field spaces. In this case, geometrical destabilisation and non-geodesic motion in field space can yield a large peak in the scalar power spectrum on small scales, and these enhanced scalar perturbations can lead to primordial black hole production and second-order gravitational wave generation.
It would be interesting to study possible constraints on very small values of $\alpha$ in multi-field $\alpha$-attractor models in a Bayesian framework.

\subsection{Future prospects}
\label{sec:future prospective}
Future, ground-based CMB-S4 experiments are designed to achieve observational constraints $\sigma(n_s)= 0.0019$ and $\sigma(r)= 10^{-3}$ \cite{CMB-S4:2016ple}. 
\begin{figure}
\centering
\includegraphics[width=0.6\textwidth]{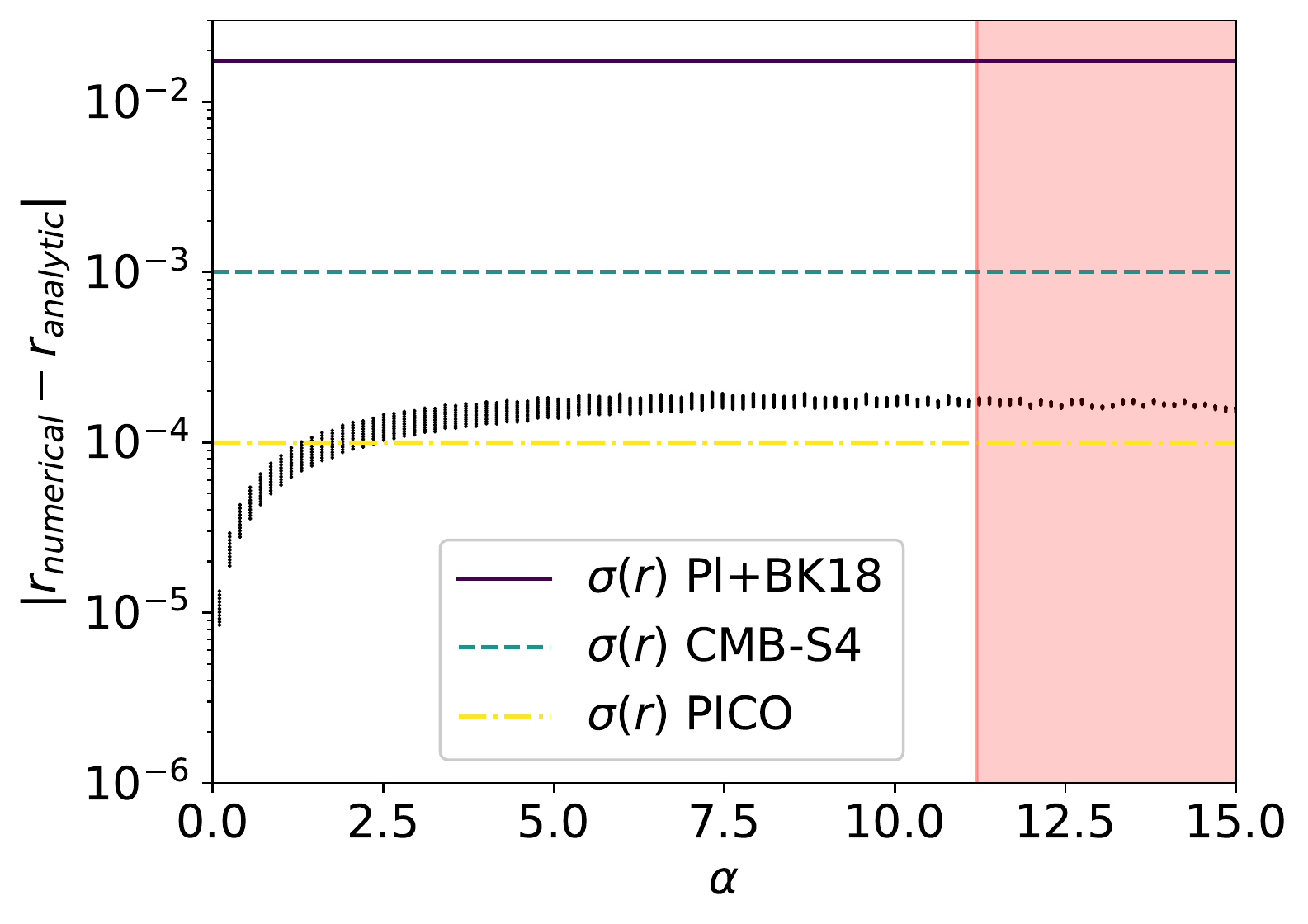}
\caption{For T-models with $p=2$ and $\alpha<15$, we represent the absolute value of the difference between the numerical $r$ results and the improved prediction \eqref{r new p dependence}. For each $\alpha$, different points corresponds to a different reheating duration, in the range $0\leq \Delta \tilde N_\text{rh}\leq \Delta \tilde N_\text{rh,max}(\alpha)$, see Figure \ref{fig:max duration of reheating}. The horizontal lines represent the values of $\sigma(r)$ for the \textit{Planck}+BICEP/Keck 2018, CMB-S4 and PICO surveys. We highlight in red $\alpha$ values that are excluded at $95\%$ C.L. by \textit{Planck} and BICEP/Keck 2018 data, see \eqref{PlBK18 lin prior bounds extended}.}
\label{fig:check predictions for future surveys}
\end{figure}
We represent in Figure \ref{fig:check predictions for future surveys} the difference between the numerical value\footnote{When comparing with future CMB surveys, designed to achieve $\sigma(r)\leq 0.001$, we need to use the second-order slow-roll predictions for the tensor-to-scalar ratio as well, see Appendix \ref{app:slow-roll observables}.}  of $r$ and the analytical expression \eqref{r new p dependence} for T-models with $p=2$ and $\alpha<15$. These results show that the improved analytic expressions for the large-scale observables\footnote{We have separately checked that the difference between the numerical value of $n_s$ and the improved prediction \eqref{ns new p dependence} is always smaller than $\sigma(n_s)$ for both CMB-S4 and PICO.}, Eqs.\eqref{ns new p dependence}-\eqref{r new p dependence}, are accurate enough to be used when comparing T-models with CMB-S4 results. When more advanced surveys come along, e.g. the proposed space-based PICO \cite{NASAPICO:2019thw} designed to achieve $\sigma(n_s)= 0.0015$ and $\sigma(r) = 10^{-4}$, the improved analytic predictions might no longer be sufficient, see Figure \ref{fig:check predictions for future surveys}. 

A detection of primordial B-modes by future CMB surveys will result in a measurement of the parameter $\alpha$ in $\alpha$-attractor models of inflation, see e.g. \cite{Kallosh:2019eeu, Kallosh:2019hzo}. Nevertheless, even in absence of B-mode detection, it is possible to constrain $\alpha$. Tighter upper bounds on $r$ will further constrain $\alpha$ and $\log_{10}{\alpha}$ from above. 
Moreover, improvements in the measurement of $n_s$ will place stronger lower bounds on $\log_{10}{\alpha}$, thanks to the $\alpha$-dependence of $n_s$, see Eq.\eqref{ns new p dependence}.
We plan to address constraints on $\alpha$-attractors from future surveys in an upcoming work. 
    
Finally we note that the $\alpha$-attractor models considered in this work are formulated assuming that the potential $V(Z,\,\bar Z)$, see Eq.\eqref{kinetic lagrangian 1}, as well as its derivatives, are not singular at the boundary of the hyperbolic disk. The potential plateau at large values of the canonical field $\phi$ is approached exponentially and models belonging to this class are therefore referred to as \textit{exponential} $\alpha$-attractors. These include the well-known E- and T-models \cite{Kallosh:2015zsa}. \textit{Polynomial} $\alpha$-attractors \cite{Kallosh:2022feu} admit potentials with a singular derivative at the boundary of the hyperbolic disk and instead approach the plateau polynomially, e.g. 
\begin{equation}
        \label{polynomial potential}
        V(\phi) = V_0 \frac{\phi^2}{\phi^2+{\phi_0}^2} \;,
\end{equation}
where the parameter $\phi_0$ describes the polynomial approach to the plateau at large field values, $V(\phi)\simeq V_0 (1-{\phi_0}^2/\phi^2+\cdots)$. These models have been considered, e.g., in primoridal black hole formation scenarios \cite{Braglia:2020eai, Kallosh:2022vha} (see \cite{Iacconi:2023slv} for the analysis of non-Gaussianity). The universal predictions derived within exponential and polynomial $\alpha$-attractors are different \cite{Kallosh:2022feu}. For example, using the model in Eq.\eqref{polynomial potential}, the universal predictions for the scalar spectral tilt and tensor-to-scalar ratio depend on $\Delta N_\text{CMB}$ and $\phi_0$. We leave investigations of polynomial attractors for future work.   

\section*{Acknowledgments}
The authors are especially grateful to Seshadri Nadathur for help with the MCMC sampler. LI would like to thank Guilherme Brando, Juan Carlos Hidalgo, Natalie B. Hogg and Francisco X. Linares Cedeño for interesting discussions. LI is supported by a Royal Society funded postdoctoral position. MF acknowledges support from the ``Ram\'{o}n y Cajal'' grant RYC2021-033786-I. MF's  work is partially supported by the Spanish Research Agency (Agencia Estatal de Investigaci\'{o}n) through the Grant IFT Centro de Excelencia Severo Ochoa No CEX2020-001007-S, funded by MCIN/AEI/10.13039/501100011033. JV acknowledges support from  Ruth och Nils-Erik Stenb{\"a}cks Stiftelse and the Academy of Finland grant 347088. DW is supported by the Science and Technology Facilities Council, grant number ST/W001225/1. 
Supporting research data are available on reasonable request from the corresponding author, Laura Iacconi.

\appendix

\section{Inflaton oscillations}
\label{app: inflaton oscillations}
At the end of inflation, the inflaton exits the slow-roll regime and starts oscillating. For $\alpha$-attractor T-models with $p=2$, the amplitude of the inflaton oscillations, $\phi_m$, depends on the value of $\alpha$ and determines the (time-averaged) equation-of-state parameter during the initial phase of the inflaton oscillations \cite{Lin:2023ugk} (see also \cite{Lin:2023jls}). 
\begin{figure}
    \centering
    \includegraphics[width = \textwidth]{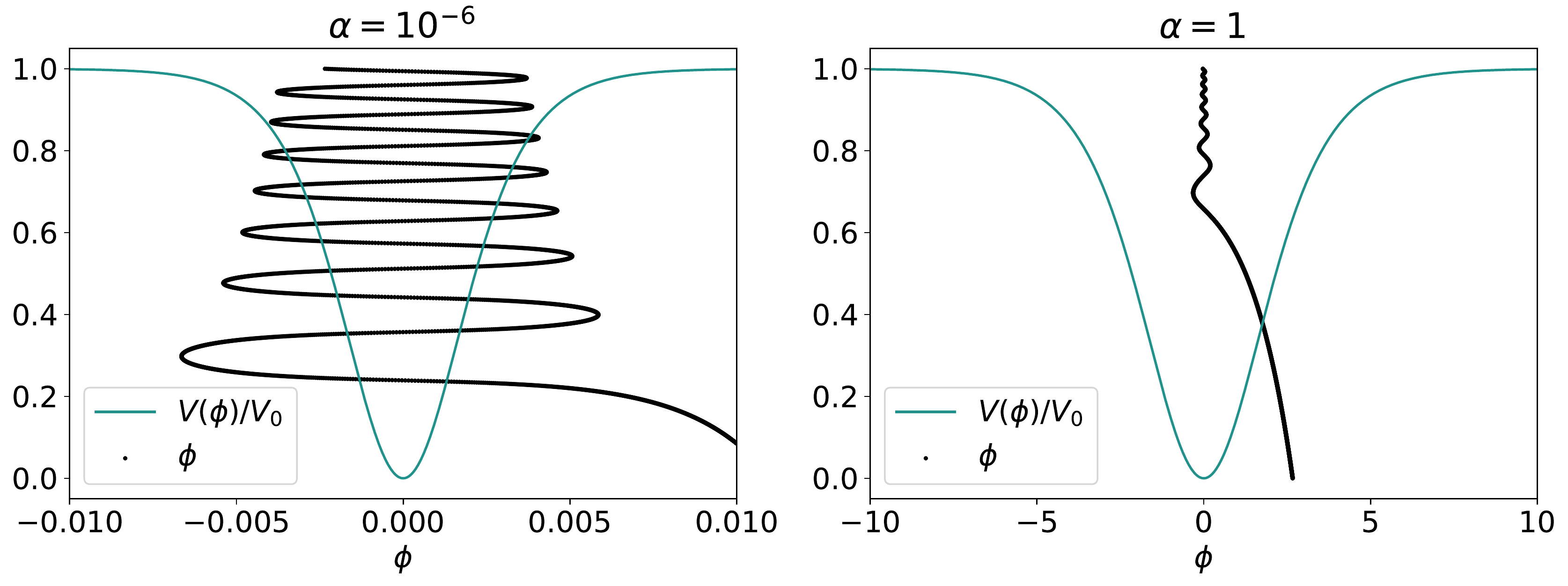}
    \caption{Profile of the normalised inflaton potential (green line) and inflaton fluctuations (black dots) for two T-models with $p=2$ and different values of $\alpha$. In order to represent the inflaton oscillations, we identify the oriented, vertical axis with an appropriately renormalised version of e-folding time. The oscillations represented take place between the times when $\epsilon_H$ crosses unity for the first and 30th times.}
    \label{fig:field oscillation amplitude}
\end{figure}

We demonstrate this concept in Figure \ref{fig:field oscillation amplitude} for two exemplary models with small and large $\alpha$ respectively. By comparing the amplitude of $\phi_m$ with the shape of the potential, we see that for the T-model with $\alpha=1$ (right panel) the inflaton 
performs simple damped oscillations about the minimum of its potential, and we can approximate the potential at the end of inflation as in Eq.~\eqref{potential small phi}, $V(\phi)\sim \left(\phi/\sqrt{6\alpha} \right)^2$, leading to an effective equation of state $w=0$.
In the left panel we see that the case with $\alpha=10^{-6}$ is rather different: the inflaton field is oscillating but, due to the initial amplitude $|\phi_m|>\sqrt{6\alpha}$, spends most of the time during its initial oscillations on the flat plateau of its potential, rather than it does in the quadratic minimum.
During this regime, $\epsilon_H$ is oscillating rapidly and, even if periodically it gives $\epsilon_H>1$, on average $\epsilon_H$ is smaller than unity, i.e. 
\begin{equation}
    \label{time averaged epsilonH}
    \langle \epsilon_H(t) \rangle = T^{-1}\int_t^{t+T}\mathrm{d}t' \, \epsilon_H(t') <1 \;, 
\end{equation}
where $T$ is the period of a single oscillation. 

This type of dynamics goes by the name of oscillating inflation \cite{Damour:1997cb, Liddle:1998pz, Taruya:1998cz, Cardenas:1999cw, Lee:1999pta, Sami:2001zd, Koutvitsky:2016rkw}
and it can extend the effective duration of inflation for small $\alpha$. 

Only when the amplitude of the inflaton oscillations is damped enough, the field oscillates only around the quadratic minimum of the potential, i.e. $\langle \epsilon_H \rangle$ becomes larger than unity, inflation ends and conventional reheating starts. Note that the solution represented in the left panel of Figure \ref{fig:field oscillation amplitude} does not reach this regime for the times plotted. Nonetheless the actual expansion, expressed in terms of e-folds, is small.

During the initial oscillating-inflation phase, the averaged equation-of-state parameter may be $\langle w \rangle \sim -1$, as demonstrated analytically in \cite{Lin:2023ugk} for T-models with $p=2$. Here we check the end-of-inflation phenomenology for models with $\alpha\geq 10^{-8}$ and numerically confirm the findings of \cite{Lin:2023ugk}. Importantly, we also estimate the duration of the oscillating-inflation phase, 
since an extended phase of oscillating inflation would impact the large-scale predictions, see Eq.~\eqref{Nstar}. 

\begin{figure}
\centering
\includegraphics[width=\textwidth]{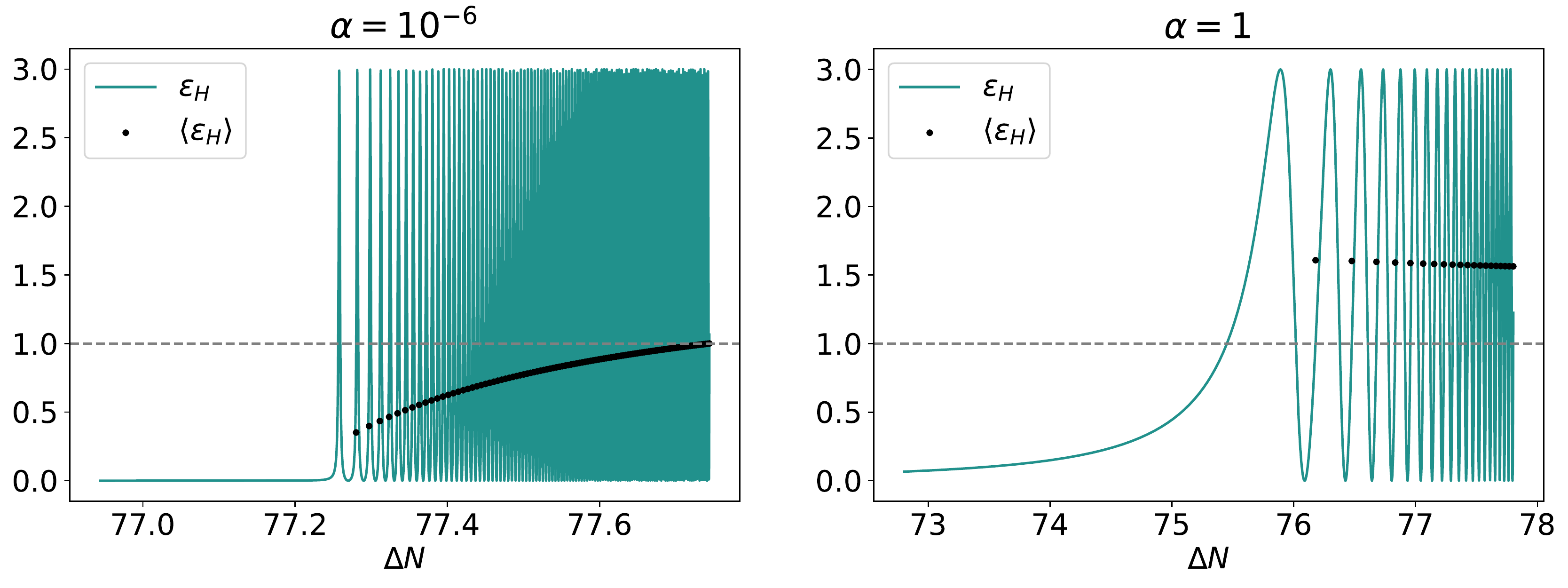}
\caption{Numerical evolution of $\epsilon_H$ (green line) and the averaged quantity $\langle \epsilon_H \rangle$ (black dots), calculated for two T-models with $p=2$ and different values of $\alpha$. We mark unity with a horizontal, dashed line. In the left panel we represent the $N$ range up until $\langle \epsilon_H \rangle$ becomes larger than unity. In the right panel the first $\langle \epsilon_H \rangle$ value is already larger than unity, and for clarity we represent a time range that allows to see many inflaton oscillations.}
\label{fig:epsilon oscillation inflation}
\end{figure}
We display in Figure \ref{fig:epsilon oscillation inflation} the time evolution of $\epsilon_H$ for our two working examples. By comparing the instantaneous time-dependence of $\epsilon_H$ with the time-averaged one\footnote{We numerically evaluate the time average of $\epsilon_H$ by integrating the solution over the period of one (first black point), two (second black point), ... full oscillations. In other words, we define the periods $T_i$, each one used to produce one of the black points, as $T_i\equiv N(\epsilon_H=1|_{(1+2i)^\text{th}\text{ time}})-N(\epsilon_H=1|_{1^\text{st}\text{ time}})$. For analytical studies  of oscillating inflation (see e.g. \cite{Cardenas:1999cw}), the period is usually defined as $T\equiv 4\int_0^{\phi_m}\mathrm{d}\phi \; [2(V(\phi_m)-V(\phi))]^{1/2}$.}, we see that for the model with small $\alpha$ it takes numerous oscillations for inflation to end, i.e. to reach the regime with $\langle \epsilon_H \rangle$ larger than unity.

\begin{figure}
\centering
\includegraphics[width=0.7\textwidth]{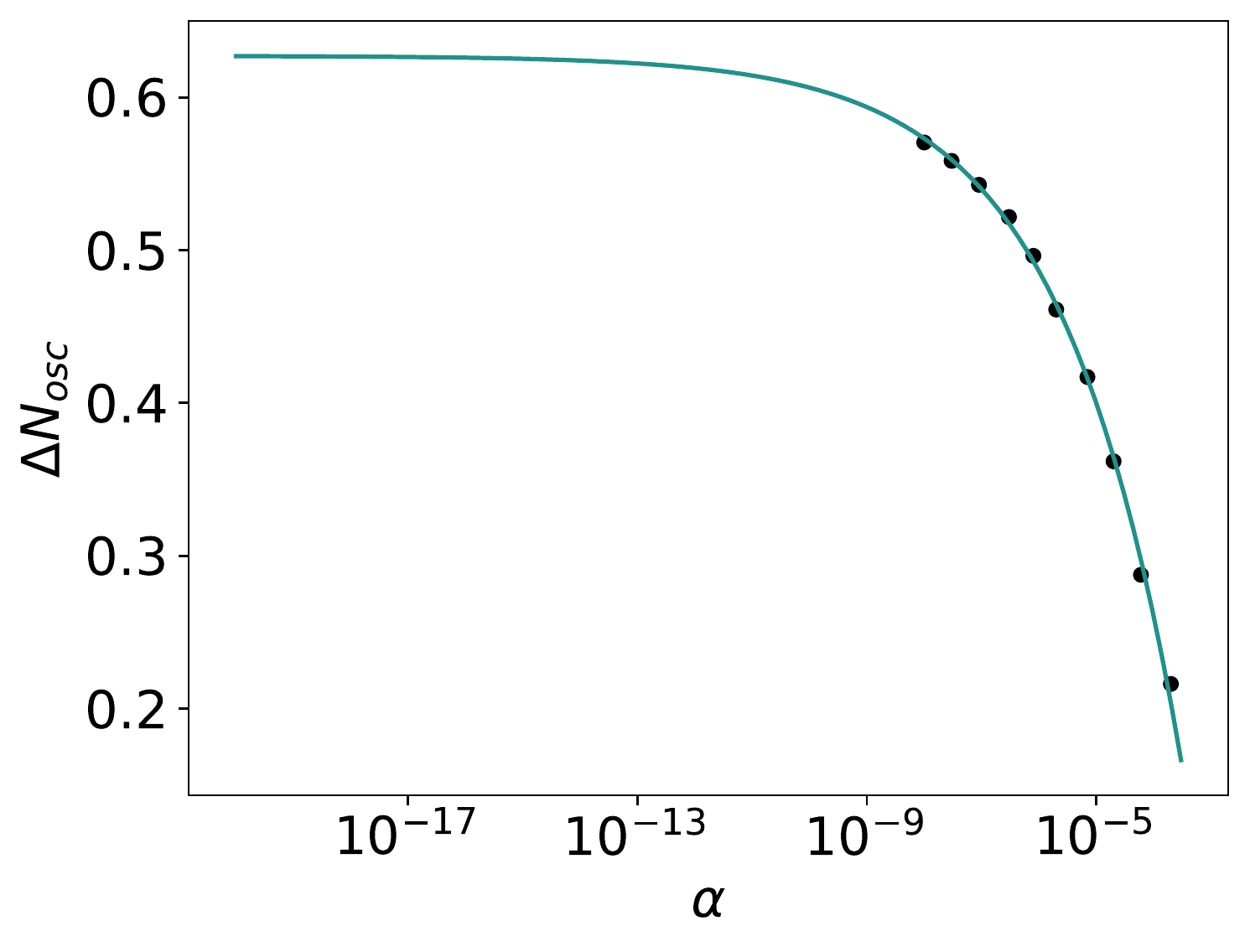}
\caption{Number of e-folds of oscillating inflation for T-models with $p=2$ and different values of $\alpha$ (black dots). The green curve represents the fitting function \eqref{fit oscillations}, which we plot for $\alpha\leq 3\times 10^{-4}$, i.e. over $\alpha$ values that give rise to a phase of oscillating inflation.}
\label{fig:duration oscillatory inflation}
\end{figure}
We show in Figure \ref{fig:duration oscillatory inflation} the duration of oscillatory inflation $\Delta N_\text{osc}$ (defined as the e-folding time elapsed between the time when $\epsilon_H$ first becomes larger than unity and the time when $\langle \epsilon_H \rangle$ become larger than one), against values of $\alpha$. We see that $\Delta N_\text{osc}$ increases for decreasing $\alpha$, but it always remains small, reaching $0.57$ e-folding for the lowest $\alpha$ considered. 

The numerical calculation of $\Delta N_\text{osc}$ is not a trivial task, as it involves integrating highly-oscillatory functions, and extending it to models with $\alpha<10^{-8}$ is extremely time-consuming. For this reason we extrapolate values of $\Delta N_\text{osc}$ to smaller $\alpha$ by using the fitting function 
\begin{equation}
    \label{fit oscillations}
    \Delta N_\text{osc}(\alpha) = 0.627193 -2.4885\, \alpha^{0.207873} \;,
\end{equation}
which shows excellent agreement with the numerical values. It implies
that even for
arbitrarily small values of $\alpha$ the phase of oscillatory inflation is small, with $\Delta N_\text{osc}< 0.627193$. For this reason we will neglect it. 

\section{Slow-roll large-scale observables}
\label{app:slow-roll observables}
In section \ref{sec:slow-roll approximation} we have described the slow-roll approximation and introduced the potential (PSRP) and Hubble (HSRP) slow-roll parameters, see \eqref{PSRP} and \eqref{HSRP phi N}. The PSRP and the HSRP can be related to each other \cite{Liddle:1994dx}, 
\begin{align}
\label{exact epsilon_V}
    \epsilon_V &= \epsilon_H \left(\frac{3-\eta_H}{3-\epsilon_H} \right)^2 \\
    \label{exact eta_V}
    \eta_V &=  \frac{\eta_H'}{3-\epsilon_H} + \frac{(3-\eta_H)(\eta_H+\epsilon_H)}{3-\epsilon_H}\;.
\end{align}
In the slow-roll approximation, the scalar spectral tilt and tensor-to-scalar ratio on large scales are given in terms of the slow-roll parameters, evaluated when the CMB scale, $k_\text{CMB}$, crossed the horizon, i.e. $\Delta N_\text{CMB}$ e-folds before the end of inflation. At second order in the slow-roll parameters the spectral tilt is \cite{Stewart:1993bc,Liddle:1994dx}
\begin{align}
\label{ns second order PSRP}
    n_s -1 &= -6\epsilon_V +2\eta_V  +4\left( \frac{11}{3}-\frac{3}{2}c \right) {\epsilon_V}^2 -2(7-2c) \epsilon_V \eta_V +\frac{2}{3} {\eta_V}^2 +\frac{1}{2}\left(\frac{13}{3}-c\right){\xi_V}^2  \\
    \label{ns second order HSRP}
    &= -4\epsilon_H +2\eta_H -2(1+c) \epsilon_H^2 +\frac{1}{2} (-3+5c) \epsilon_H \eta_H +\frac{1}{2}(3-c){\xi_H}^2 \;,
\end{align}
where $n_s$ is expressed in terms of the PSRP, see \eqref{PSRP}, and HSRP, see \eqref{HSRP phi N}, in the first and second lines respectively. Additionally, $c=4(\log{(2)}+\gamma) -5$, where $\gamma$ is Euler-Mascheroni constant. The first-order results are given by the first two terms in both lines, i.e. $n_s-1= -6\epsilon_V +2\eta_V$ and $n_s-1= -4\epsilon_H +2\eta_H$ using the PSRP and the HSRP respectively. 
\begin{figure}
\centering
\includegraphics[width = \textwidth]{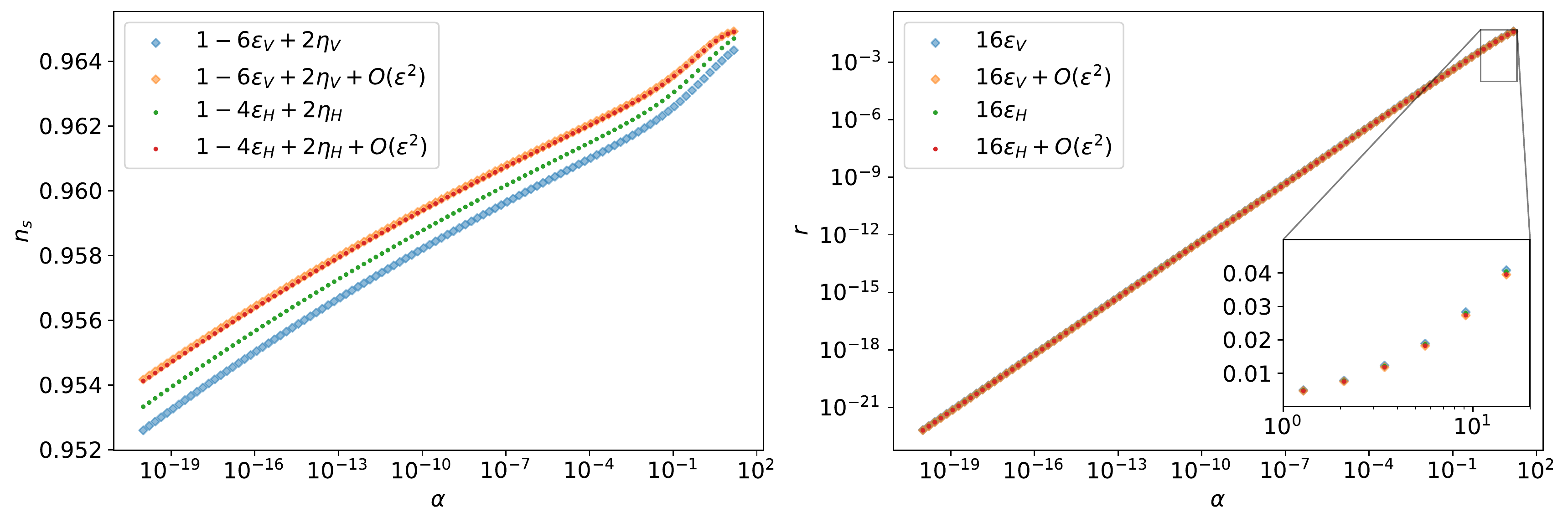}
\caption{First- and second-order slow-roll $n_s$ (left panel) and $r$ (right panel) predictions for $\alpha$-attractors T-models with potential \eqref{single field potential}, $p=2$ and $10^{-20}\leq \alpha \leq 15$. The results are obtained assuming instant reheating. In the left panel the two series of points corresponding to the second-order HSRP (red) and PSRP (orange) results are almost overlapping. On the right this is the case for both the first- and second-order predictions for $r$.}
\label{fig:ns r first and second order}
\end{figure}
In the left panel of Figure \ref{fig:ns r first and second order} we show the first- and second-order slow-roll $n_s$ results for $\alpha$-attractor T-models with $p=2$ and $10^{-20}\leq \alpha\leq 15$. These results are obtained by substituting the numerical background solutions into the expressions \eqref{ns second order PSRP} and \eqref{ns second order HSRP}\footnote{For the results employing the HSRP, we use the Klein--Gordon equation \eqref{single field inflaton eom}, written in terms of $N$, and its derivative to calculate $\phi''(N)$ and $\phi'''(N)$ respectively.}. We use the numerical values of $\Delta N_\text{CMB}$, calculated for each model by iteratively solving Eq.~\eqref{Nstar} with values of $V_0$ compatible with CMB measurements. 

The predictions in Figure \ref{fig:ns r first and second order} show a logarithmic dependence on $\alpha$, similar to that obtained when plotting the universal prediction expression \eqref{ns universal} in the left panel of Figure \ref{fig:ns and r universal predictions}. The first-order results calculated with the PSRP and the HSRP differ by $10^{-3}$ at most in the $\alpha$ range represented, while the second-order results agree at this level of precision, $|n_{s\,\text{HSR,so}}-n_{s\, \text{PSR,so}}|\lesssim 9\times 10^{-5}$, and are practically indistinguishable. Since the dependence of $n_s$ on $\alpha$ is logarithmic, a small error in determining $n_s$ would potentially result into an order-of-magnitude error in $\alpha$. For this reason, in section \ref{sec:improved predictions} we decide to use the second-order slow-roll predictions for $n_s$. 

The tensor-to-scalar ratio at second order in the slow-roll parameters is \cite{Stewart:1993bc, Martin:2006rs}
\begin{align}
\label{r PSRP}
r & = 16\epsilon_V \left[
1+\left(\frac{c}{2}-\frac{13}{6}\right)\left( 2\epsilon_V-\eta_V \right)  \right] \\
\label{r HSRP}    
& = 16\epsilon_H \left[1 + \left(\frac{c}{2} -\frac{3}{2} \right)(\epsilon_H-\eta_H ) \right] \;,
\end{align}
where in the first (second) line the PSRP (HSRP) are used. The first-order results is given by the first term in each line. We show in the right panel of Figure \ref{fig:ns r first and second order} the numerical results obtained at first- and second-order by employing Eqs.~\eqref{r PSRP} and \eqref{r HSRP}. The value of $r$ depends linearly on $\alpha$, in accordance with the qualitative behavior represented in the right panel of Figure \ref{fig:ns and r universal predictions}, using the universal prediction \eqref{r universal} and Eq.~\eqref{N CMB}. Notably, the first- and second-order results are practically indistinguishable. In particular, $|r_\text{HSR,fo}-r_\text{PSR,fo}|\lesssim 7\times 10^{-4}$, $|r_\text{HSR,so}-r_\text{PSR,so}|\lesssim 5 \times 10^{-5}$,  $|r_\text{PSR,fo}-r_\text{PSR,so}|\lesssim 3\times 10^{-3}$ and $|r_\text{HSR,fo}-r_\text{HSR,so}|\lesssim 2 \times 10^{-3}$. For this reason and for simplicity, in section \ref{sec:improved predictions} we choose to use the first-order results for $r$, and in particular we will use the HSRP expression, 
\begin{equation}
\label{r first order HSRP}
    r=16\epsilon_H \;.
\end{equation}

\bibliography{refs} 
\bibliographystyle{JHEP}

\end{document}